\documentclass[acmsmall,screen]{acmart}
\usepackage{amsmath}
\usepackage{algorithm}
\usepackage{algorithmic}

\usepackage{graphicx}
\usepackage{color}
\usepackage[export]{adjustbox}
\usepackage{subfigure}
\usepackage{booktabs}
\usepackage{multirow}
\usepackage{array}
\usepackage{enumitem}
\usepackage{url}
\usepackage{amsmath}
\usepackage{amsthm}
\usepackage{setspace}
\usepackage{nicefrac}

\usepackage{pdftexcmds}
\usepackage{catchfile}
\usepackage{ifluatex}
\usepackage{ifplatform}

\ifwindows
\usepackage{times}
\usepackage{lineno}
\linenumbers
\usepackage{citesort}
\fi
\iflinux
\fi
\ifmacosx
\fi 

\graphicspath{ {img/} }

\newtheorem{theorem}{Theorem}
\newtheorem{lemma}[theorem]{Lemma}
\theoremstyle{definition}
\newtheorem{definition}{Definition}

\def\tC{\textbf{C}}
\def\tW{\textbf{W}}
\def\bR{\mathbf{R}}

\def\bW{\mathbf{W}}
\def\bu{\mathbf{u}}
\def\bb{\mathbf{b}}
\def\bh{\mathbf{h}}
\def\be{\mathbf{e}}
\def\ba{\mathbf{a}}
\def\bs{\mathbf{s}}
\def\bz{\mathbf{z}}
\def\bq{\mathbf{q}}

\newcommand{\NM}[2]{\left\|  #1 \right\|_{#2}}
\newcommand{\Px}[2]{\text{prox}_{#1}\left( #2 \right) }
\newcommand{\SO}[1]{P_{\mathbf{\Omega}}\left( #1 \right)}

\usepackage{array}
\newcolumntype{L}[1]{>{\raggedright\let\newline\\\arraybackslash\hspace{0pt}}m{#1}}
\newcolumntype{C}[1]{>{\centering\let\newline  \\\arraybackslash\hspace{0pt}}m{#1}}
\newcolumntype{R}[1]{>{\raggedleft\let\newline \\\arraybackslash\hspace{0pt}}m{#1}}

\newtheorem{remark}{Remark}[section]

\begin{document}



\title{Side Information Fusion for Recommender Systems over Heterogeneous Information Network}

\author{Huan Zhao}
\affiliation{\institution{4Paradigm Inc., Shenzhen China}}
\affiliation{
	\institution{Department of Computer Science and Engineering,
		Hong Kong University of Science and Technology, Hong Kong}
}
\email{hzhaoaf@cse.ust.hk}
\author{Quanming Yao}
\affiliation{\institution{4Paradigm Inc., Hong Kong}}
\affiliation{\institution{Department of Electronic Engineering, Tsinghua University, Beijing, China}}
	\email{qyaoaa@cse.ust.hk}
\author{Yangqiu Song}
\affiliation{%
	\institution{Department of Computer Science and Engineering,
		Hong Kong University of Science and Technology, Hong Kong}
}
\email{yqsong@cse.ust.hk}
\author{James T. Kwok}
\affiliation{%
	\institution{Department of Computer Science and Engineering,
		Hong Kong University of Science and Technology, Hong Kong}
}
\email{jamesk@cse.ust.hk}
\author{Dik Lun Lee}
\affiliation{%
	\institution{Department of Computer Science and Engineering,
		Hong Kong University of Science and Technology, Hong Kong}
}
\email{dlee@cse.ust.hk}

\renewcommand\shortauthors{Zhao, H. et al}

\begin{abstract}
Collaborative filtering (CF) has been one of the most important and popular recommendation methods, which aims at predicting users' preferences (ratings) based on their past behaviors. Recently, various types of side information beyond the explicit ratings users give to items, such as social connections among users and metadata of items, have been introduced into CF and shown to be useful for improving recommendation performance. 
However, previous works process different types of information separately, thus failing to capture the correlations that might exist across them.
To address this problem, in this work, we study the application of heterogeneous information network (HIN), 
which offers a unifying and flexible representation of different types of side information, 
to enhance CF-based recommendation methods. 
However, we face challenging issues in HIN-based recommendation, i.e., how to capture similarities of complex semantics between users and items in a HIN, and how to effectively fuse these similarities to improve final recommendation performance.
To address these issues, we apply metagraph to similarity computation and solve the information fusion problem with a ``matrix factorization (MF) + factorization machine (FM)'' framework. For the MF part, we obtain the user-item similarity matrix from each metagraph and then apply low-rank matrix approximation to obtain latent features for both users and items.
For the FM part, we apply FM with Group lasso (FMG) on the features obtained from the MF part to train the recommending model and, 
at the same time, identify the useful metagraphs. Besides FMG, a two-stage method, we further propose an end-to-end method, hierarchical attention fusing (HAF), to fuse metagraph based similarities for the final recommendation.
Experimental results on four large real-world datasets show that the two proposed frameworks significantly outperform existing state-of-the-art methods in terms of recommendation performance.
\end{abstract}

\keywords{Recommender Systems, Collaborative Filtering, Heterogeneous Information Networks, Matrix Factorization, Factorization Machine, Graph Attention Networks}

\maketitle

\section{Introduction}
\label{sec-intro}

In the big data era, people are overwhelmed by the huge amount of information on the Internet, making recommender systems (RSs) an indispensable tool for getting interesting information.  
Collaborative filtering (CF) has been the most popular recommendation method in the last decade~\cite{herlocker1999algorithmic,koren2008factorization}, which tries to predict a user's preferences based on their past behaviors.
In recent years, researchers try to incorporate auxiliary information beyond users' behaviors, or \textit{side information}, to enhance CF. 
For example, social connections among users~\cite{ma2011recommender,zhao2017sloma,Xiao2019Sean}, 
reviews of items~\cite{mcauley2013hidden,ling2014ratings}, metadata attached to commodities~\cite{Wang2018Billion},  or locations of users and items~\cite{ye2011exploiting,zheng2012towards},
have been shown to be effective for improving recommendation performance. 
However, a major limitation of most existing methods is that various types of side information are processed independently, leading to information loss across different types of side information. 
This limitation becomes more and more severe, because
modern websites record rich side information about their users and contents~\cite{pan2016survey} and it would be a huge loss to their business if the side information cannot be fully utilized to improve performance.
For example, on Yelp~(\url{https://www.yelp.com/}), a website recommendation business to users,  users can follow other users to form a social network, businesses have categories and locations, and users can write reviews on businesses. If each type of side information is processed in isolation, information that exists across different types of side information will be neglected. Therefore, a unifying framework is needed to fuse all side information for producing effective recommendations.

\begin{figure}[t]
	\centering
	\includegraphics[width = 0.8\textwidth]{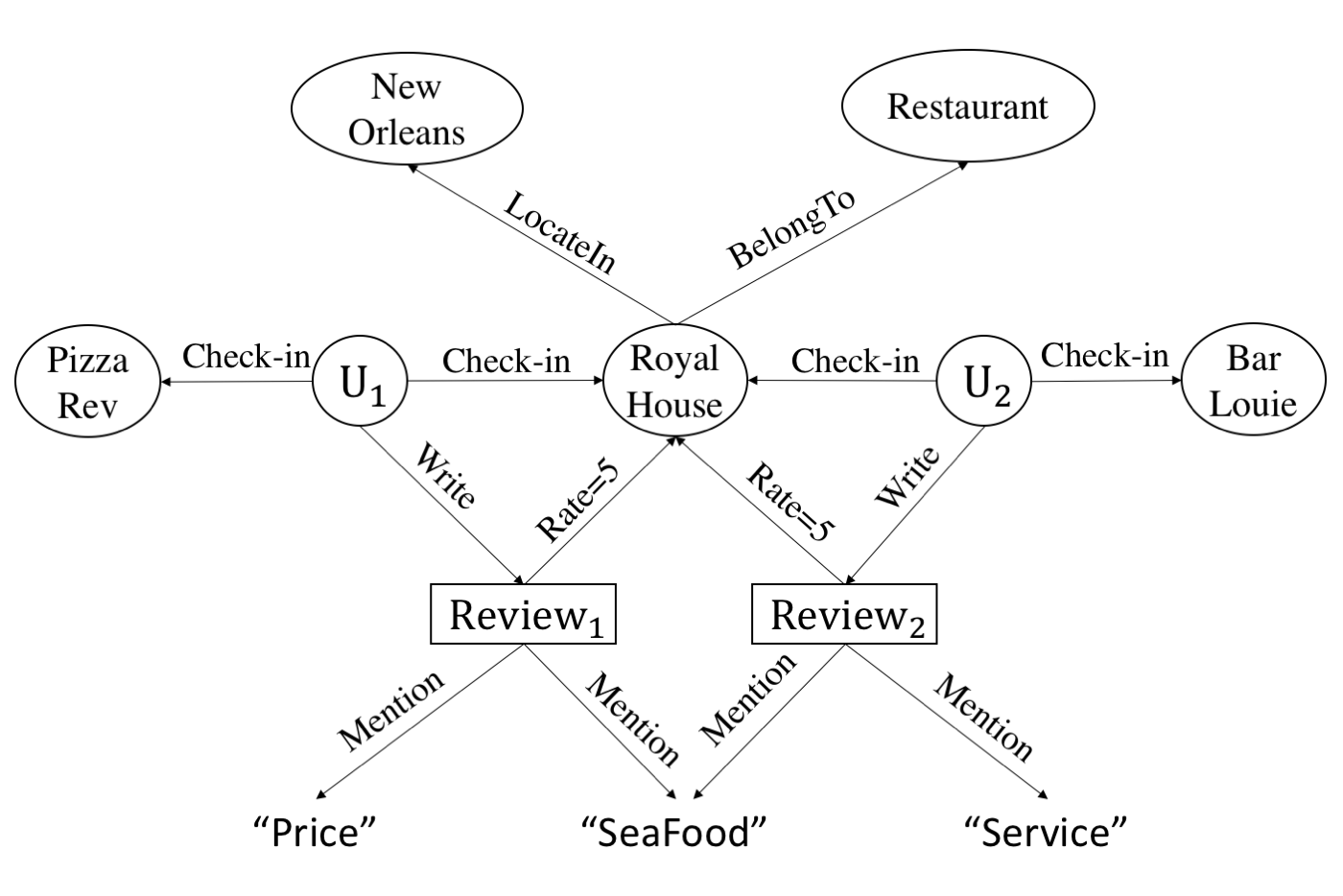}
	\vspace{-10px}
	\caption{An example HIN built for Royal House on Yelp. We can see that various side information, like Category (Restaurant) and City (New Orleans), are modeled as nodes (entities) of different types. And the corresponding relations among different side information are modeled by edges. }
	\label{fig-intro-example}
\end{figure}

\begin{figure}[ht]
\centering
\includegraphics[width=0.65\textwidth]{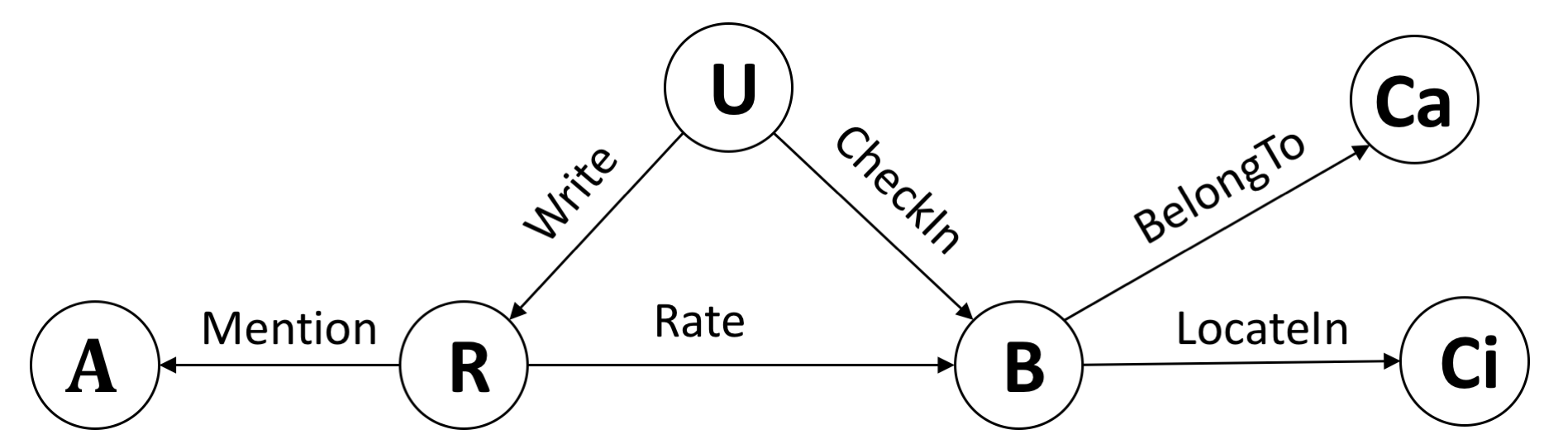}
\vspace{-10px}
\caption{The network schema for the HIN in Figure~\ref{fig-intro-example}. A: aspect extracted from reviews; R: reviews; U: users; B: business; Cat: category of business; Ci: city.}
\label{fig-hin-schema}
\end{figure}

Heterogeneous information networks (HINs)~\cite{sun2011pathsim,shi2017survey} have been proposed as a general data representation tool for different types of information, such as scholar network~\cite{sun2011pathsim} and knowledge graph~\cite{wang2015incorporating}. Thus, by modeling various side information and the relations between them as different nodes (entities) and edges, respectively, HIN can be used as a unifying framework for RSs \cite{yu2014personalized,shi2015semantic}. Figure~\ref{fig-intro-example} shows an example HIN on Yelp, and Figure~\ref{fig-hin-schema} shows a network schema defined over the entity types {\it User}, {\it Review},  {\it Aspect}, {\it Business}, etc. 
Based on the network schema, we can design metapaths~\cite{sun2011pathsim,shi2017survey}, which are sequences of node types, to compute the similarities between users and businesses for generating recommendations.
For example, we can define complicated metapaths such as ${\it User}\rightarrow {\it Review}\rightarrow {\it Aspect}\rightarrow {\it Review}\rightarrow {\it User} \rightarrow {\it Business}$, to measure similarities between user and business based on similar reviews written by users about the same aspect. Therefore, we can unify different side information with HIN and design metapaths to compute user-item similarities of different semantics for making more effective recommendation.

However, there are \textbf{two major issues} facing existing HIN-based RSs.
The first issue is that metapaths are not enough for representing the rich semantics for HIN-based RSs. We refer to it as \textit{semantic limitation}.
Figure~\ref{fig-intro-example} shows a concrete example, where the metapath ${\it User}\rightarrow {\it Review}\rightarrow {\it Aspect}\rightarrow {\it Review}\rightarrow {\it User}\rightarrow {\it Business}$ is used to capture users' similarity since both users write reviews and mention the same aspect (seafood) in the review texts. 
However, if we want to capture the similarity induced by the two users' reviews mentioning the same aspect (such as \textit{seafood}) and ratings on the same business (such as \textit{Royal House}), then metapath is not able to capture this semantic. Thus, we need a better way to capture such complicated semantics. Recently, 
\citet{Huang2016MetaSC} and \citet{Fang2016SemanticPS,fang2019metagraph} proposed to use metagraph (or meta-structure) for computing similarity between homogeneous types of entities (e.g., using {\it Person} to search  {\it Person}) over HINs, which is more powerful than metapath in capturing complex semantics. However, they did not explore metagraphs for entities of heterogeneous types, which are essential for RSs. In this paper, we extend metagraph to capture similarities of complex semantics between users and items (businesses) in recommendation scenarios.

The second issue is about \textit{similarity fusion}, i.e., how to fuse the similarities of different semantics between users and items for recommendation. Our goal is to achieve accurate predictions of the users' ratings given these similarities.
In general, there are three groups of approaches to do so.
The first ones are directly using the metapath based similarity to improve the rating prediction task. For example, in \cite{shi2015semantic}, the proposed method firstly computes multiple similarity matrices between users and items based on different metapaths, and then learn a weighted mechanism to explicitly combine these similarities to approximate the user-item rating matrix.
However, these similarity matrices could be too sparse to contribute to the final ensemble.
The second ones are to firstly factorize each user-item similarity matrix to obtain user and item latent features, and then use all latent features to recover the user-item rating matrix \cite{yu2014personalized,shi2018heterogeneous}. These methods solve the sparsity problem associated with each similarity matrix. However, it does not fully utilize the latent features because they ignore the latent feature interactions among different metapaths and only capture linear interactions among the latent features. 
The third group of approaches are similar to the second ones, while they propose to use neural network based methods to extract and fuse latent features in an end-to-end manner~\cite{Hu2018LMB,fan2019metapath,jin2020efficient}. 
However, all these methods still rely on metapath, thus failing to capture complex semantics. In one words, existing HIN-based recommendation methods~\cite{yu2014personalized,shi2015semantic,Hu2018LMB,fan2019metapath,jin2020efficient} suffer from information loss in various ways.

To address the above challenges, we propose a new systematic way to fuse various side information in HIN for recommendation. 
First, instead of using metapath for recommendation~\cite{yu2014personalized,shi2015semantic}, we introduce the concept of metagraph to the recommendation problem, which allows us to incorporate more complex semantics into HIN-based RSs.
Second, instead of computing the recovered matrices directly from the metagraphs, we utilize the latent features from all metagraphs. Based on matrix factorization (MF)~\cite{mnih2007probabilistic,koren2008factorization} and factorization machine (FM)~\cite{rendle2012fm}, we propose a ``MF + FM'' framework for our metagraph based RS in HIN.
We first compute the user-item similarity matrix from each metagraph and then factorize the similarity matrix to obtain a set of vectors representing the latent features of users and items. 
Finally,  we use FM to assemble these latent features to predict the missing ratings that users give to the items. This method enables us to capture nonlinear interactions among all latent features, which has been demonstrated to be effective in FM-based RS~\cite{rendle2012fm}.
To further improve the performance of the ``MF+FM'' framework, we propose to use group lasso \cite{jacob2009group} with FM (denote as FMG) to learn the parameters for selecting metagraphs that contribute most to the recommendation performance. As a result, we can automatically determine which metagraphs are the most effective, and for each group of user and item features from a metagraph, how the features should be weighed. 
Besides FMG, a two-stage method to extract and fuse latent features from multiple metagraph based user-item similarities, we further propose a novel method based on graph neural networks (GNN) \cite{battaglia2018relational}, 
i.e., the hierarchical attention fusion (HAF) framework, which can fuse multiple metagraph based user-item similarities in an end-to-end manner.
Experimental results on four large real-world datasets, Amazon and Yelp, show that our frameworks significantly outperform recommendation methods that are solely based on MF, FM, or metapath in HIN. 
Our code is available at \url{https://github.com/HKUST-KnowComp/FMG}.

Preliminary results of this manuscript have been reported in KDD 2017 \cite{zhao2017meta}, 
where MF is designed to extract latent features from metagraphs, and then FMG is proposed for metagraph fusion and selection. 
In this manuscript, we formally propose and highlight the systematic framework ``MF + FM'' (see Figure~\ref{fig-framework}), which provides a complete solution for HIN-based RSs. Compared to the KDD version, we update the three key components of FMG by designing a novel latent feature extraction method for the MF part, a novel type of group lasso regularizer, and a more efficient optimization solver for the FM part, respectively. Moreover, we further propose the HAF framework (see Figure~\ref{fig-haf}), which can be regarded as a deep learning version of ``MF + FM'' pipeline, benefited from the advantages of neural networks and end-to-end training. Finally, additional experiments are performed to support the newly introduced components.

\subsection*{Notation}
We denote vectors and matrices by lowercase and uppercase boldface letters, respectively. 
In this paper, a vector always denote row vector. For a vector $\mathbf{x}$,
$\NM{\mathbf{x}}{2} = (\sum_{i = 1} |\mathbf{x}_i|^2)^{\frac{1}{2}}$ is its $\ell_2$-norm.
For a matrix $\mathbf{X}$,
its nuclear norm is
$\NM{\mathbf{X}}{*} = \sum_{i} \sigma_i(\mathbf{X})$, where $\sigma_i(\mathbf{X})$'s are 
the singular values of
$\mathbf{X}$;
$\NM{\mathbf{X}}{F} = ( \sum_{i,j}\mathbf{X}_{ij}^2 )^{\frac{1}{2}}$ is its Frobenius norm
and $\NM{\mathbf{X}}{1} = \sum_{i,j} |\mathbf{X}_{ij}|$ is its $\ell_1$-norm.
For two matrices $\mathbf{X}$ and $\mathbf{Y}$, $\langle \mathbf{X}, \mathbf{Y} \rangle = \sum_{i,j}
\mathbf{X}_{ij} \mathbf{Y}_{ij}$, and $\left[ \mathbf{X} \odot \mathbf{Y} \right]_{ij} = \mathbf{X}_{ij} \mathbf{Y}_{ij}$ denotes the element-wise multiplication.
For a smooth function $f$, $\nabla f(\mathbf{x})$ is its gradient at $\mathbf{x}$.

\section{Related Work}
\label{related}
In this section, we review existing works related to HIN, RS with various side information, and RS over HIN.

\subsection{Heterogeneous Information Networks (HINs)}
\label{sec:rel-hin}

HINs have been proposed as a general representation for many multi-typed objects and multi-relational interactions among these objects, which are often modeled as graphs or networks~\cite{sun2011pathsim,KongZY13,sun2013mining,dong2017metapath2vec,shi2017survey,yang2020heterogeneous,dong2020heterogeneous}. To capture rich semantics underlying HINs, metapath, a sequence of entity types defined by the HIN network schema, has played a key role in modeling similarities between nodes.
Based on metapath, several similarity measures, such as {PathCount}~\cite{sun2011pathsim}, {PathSim}~\cite{sun2011pathsim}, and PCRW~\cite{lao2010relational} have been proposed, and research has shown that they are useful for entity search and as similarity measure in many real-world networks. After the development of metapath, many data mining tasks have been enabled or enhanced, including recommendation~\cite{yu2013collaborative,yu2014personalized,shi2015semantic,Hu2018LMB,fan2019metapath,jin2020efficient}, similarity search~\cite{sun2011pathsim,shi2014hetesim}, clustering~\cite{wang2015incorporating,sun2013pathselclus}, classification~\cite{kong2013multi,wang2015knowsim,wang2017distant,jiang2017semihin,he2019hetespaceywalk}, link prediction~\cite{sun2012will,zhang2014meta}, malware detection~\cite{hou2017hindroid,fan2018gotcha}, and opioid user detection~\cite{fan2018automatic}. Recently, metagraph (or meta-structure) has been proposed for capturing complicated semantics in HIN that metapath cannot handle~\cite{Huang2016MetaSC,Fang2016SemanticPS,zhang2018metagraph2vec,yang2018meta,fang2019metagraph,zhao2019motif,zhang2020mg2vec}.

%

\subsection{Recommendation with Side Information}
\label{sec-rel-hin-rs}

Modern recommender systems are able to capture rich side information such as social connections among users and metadata and reviews associated with items. 
Previous works have explored different methods to incorporate heterogeneous side information to enhance CF based recommender systems. 
For example, 
\citet{ma2011recommender} and \citet{zhao2017sloma}, respectively, incorporate social relations into low-rank and local low-rank matrix factorization to improve the recommendation performance, and heterogeneous item relations are explored for recommendation in~\cite{kang2018recommendation}. 
\citet{ye2011exploiting} proposed a probabilistic model to incorporate users' preferences, social network and geographical information to enhance point-of-interests recommendation. In~\cite{wang2019exploring}, knowledge graph is used to enhance the item representation in textual content recommendation. In~\cite{Xiao2019Sean,Xiao2020Sean}, social connections and textual features are processed together in a graph attention network framework for content recommendation. 
\citet{zheng2012towards} proposed to integrate users' location data with historical data to improve the performance of point-of-interest recommendation. \citet{Wang2018Billion,pfadler2020billion} proposed embedding based methods to incorporate side informations of commodities to improve the recommendation performance of e-commerce systems. 
These previous approaches have demonstrated the importance and effectiveness of heterogeneous side information in improving recommendation accuracy. However, most of these approaches deal with these side information separately, hence losing important information that might exist across them.

\subsection{Recommendation over Heterogeneous Information Networks}

HIN-based recommendation has been proposed to avoid the disparate treatment of different types of information. 
Based on metapath, user-item similarities of different semantics can be captured, based on which extensive methods have been proposed. In general, these methods can be roughly categorized into three groups. The first ones are to use metapath based similarities \textit{explicitly}, either as supervision signals or features. In~\cite{yu2013collaborative}, metapath based similarities are used as regularization terms in matrix factorization. In~\cite{shi2015semantic}, a linear combination prediction method is used to fuse multiple metapath based similarities for rating prediction.
The second group of methods use metapath based similarities \textit{implicitly}, 
which firstly obtain latent features of users and items from these user-item similarities, and different prediction models are designed on user and item latent features. In~\cite{yu2014personalized}, multiple metapaths are used to learn user and item latent features, which are then used to recover similarity matrices combined by a weighted mechanism.  In the KDD version of this manuscript~\cite{zhao2017meta}, our FMG framework falls into this group, where MF is used to extract latent features, while FM is used to predict by these latent features. In~\cite{shi2018heterogeneous}, metapath based embedding is utilized firstly and several fusion functions are designed to fuse these metapath based embeddings, i.e., latent features of users and items. The third group of methods are to employ recently popular neural models to fuse metapath based features or similarities. In \cite{shi2019deep}, a neural attention model is designed to extract and fuse latent features from metapath based similarities between users and items for the top-N recommendation. In \cite{Hu2018LMB}, metapath instances are treated as context for recommendation, and a co-attention model is proposed to extract and fuse user, item, and metapath contexts for the final recommendation. In \cite{fan2019metapath}, metapath is used to select neighbors, which are incorporated into the popular message passing GNN models for e-commerce recommendation. Further, in \cite{jin2020efficient}, metapath based neighborhood is defined formally, and an efficient neighborhood interaction neural model is designed to fuse user-item interactions in a more fine-grained way.

However, all these three groups of methods rely on metapath for HIN-based RS, except our FMG, thus failing to capture the complex semantics by metagraph. Besides the ``MF + FM'' framework, we further design the HAF framework, which falls in the third group, to perform the fusion and selection for all metagraph based features in an end-to-end manner. Note that despite the end-to-end training nature, conceptually, the third group of methods can still be seen as a special case of the designed pipeline in Figure~\ref{fig-framework}, since they all follow an embedding (feature extraction) $+$ multilayer perceptron (MLP) (feature fusion) pipeline, which is a standard practice in existing neural models nowadays.

\section{``MF + FM'' Framework}
\label{sec-framework}

\begin{figure}[ht]
	\centering
	\includegraphics[width=0.75\textwidth]{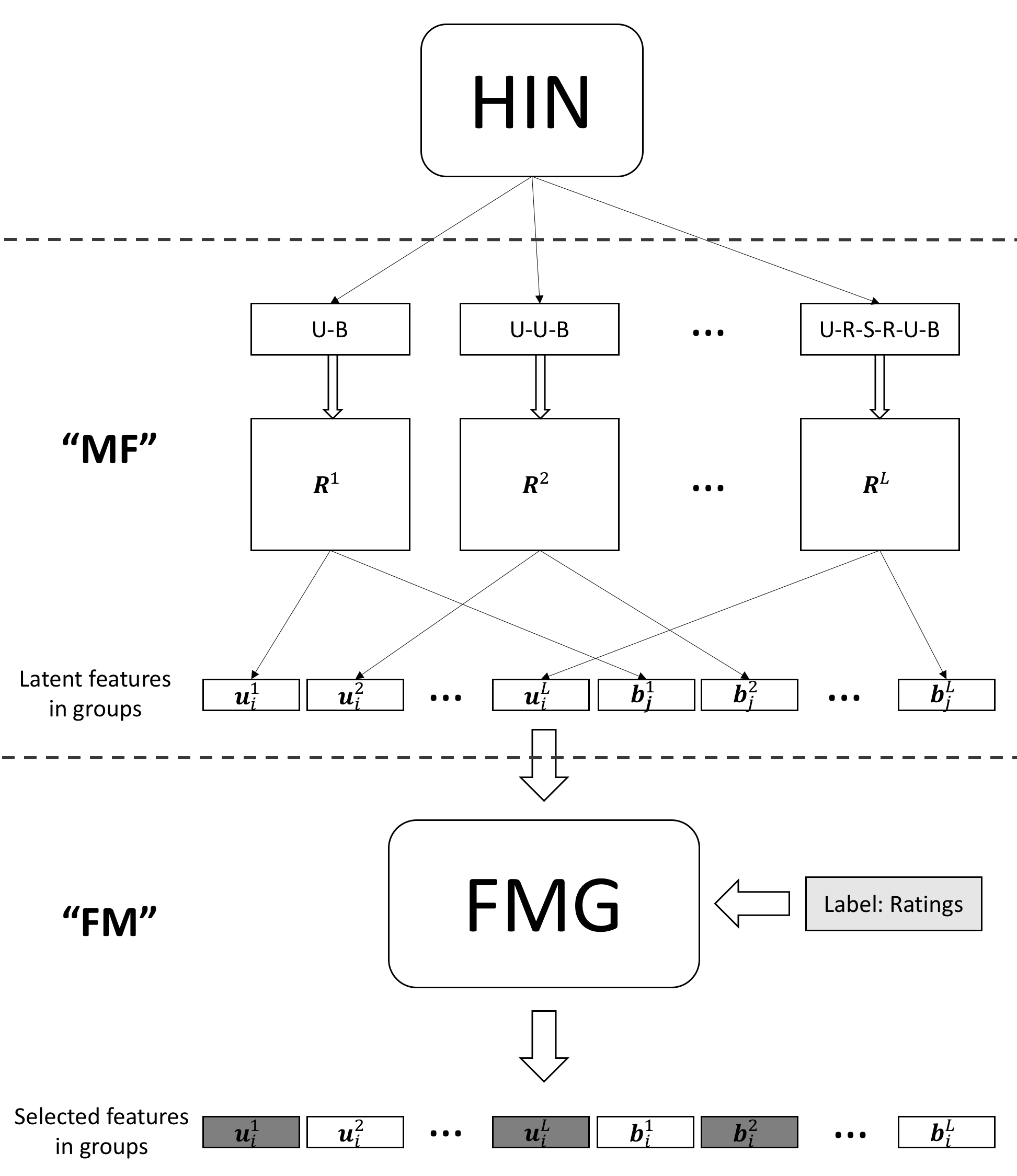}
	\vspace{-10px}
	\caption{The proposed ``MF + FM'' framework. In the MF part,
		latent features are extracted from user-item similarity matrices derived from metagraphs in a HIN (e.g., Figure~\ref{fig-intro-example}). 
		In the FM part, latent features are concatenated and then fed into FMG to predict missing ratings. 
		In the bottom, the features in grey are selected by group lasso regularizers.}
	\label{fig-framework}
\end{figure}

The proposed ``MF + FM'' framework is illustrated in Figure~\ref{fig-framework}. 
The input to the MF part is a HIN, e.g., the one in Figure~\ref{fig-intro-example}. To solve the semantic limitation issue, we propose to use metagraph instead of metapath to capture complex semantics that exists between users and items in a HIN, e.g., those in Figure~\ref{fig-yelp-metagraph} and \ref{fig-amazon-metagraph}.
Let there be $L$ metagraphs. The MF part, introduced in Section~\ref{sec-mf-part}, computes from the $L$ metagraphs $L$ user-item similarity matrices, denoted by $\bR^1, \bR^2, \cdots, \bR^L$.
We then apply low-rank matrix approximation to factorize each similarity matrix into two low-dimensional matrices, representing the latent features of users and items, which is also a classical method in CF based recommendation \cite{koren2008factorization}.
The output of the MF part is the $L$ groups of latent features for users and items. Since existing methods only compute metapath based similarities, we design a new algorithm to compute metagraph based user-item similarities.

The objective of the FM part is to utilize the latent features to learn a recommendation model that is more effective than previous HIN-based RSs. 
This addresses the similarity fusion issue.
FMG (see Section~\ref{sec-framework-model}) has two advantages over previous methods: 1) FM can capture non-linear interactions among features~\cite{rendle2012fm}, and thus more effective than linear ensemble model adopted in previous HIN-based RS~\cite{yu2014personalized}, 2) by introducing group lasso regularization, we can automatically select the useful features and in turn the useful metagraphs for a recommendation application, avoiding laborious feature and metagraph engineering when a new HIN is encountered.
Specifically, for a user-item pair, user $u_i$ and item $b_j$,
we first concatenate the latent features $\bu^1_i,\bu^2_i,\cdots,\bu^L_i$ and $\bb^1_i,\bb^2_i,\cdots,\bb^L_i$ from all of the metagraphs to create a feature vector, using rating $\bR_{ij}$ as label. We then train our FMG model with group lasso regularization method to select the useful features in the groups, where each group corresponds to one metagraph. The selected features are in grey in Figure~\ref{fig-framework}.
Finally, to efficiently train FMG, we propose two algorithms, one is based on the proximal gradient algorithm \cite{parikh2014proximal} and the other on the stochastic variance reduced gradient algorithm~\cite{xiao2014proximal} (see Section~\ref{sec:opt}). 

Note that FMG follows a two-stage pipeline, where the feature extraction and fusion are trained separately. 
Motivated by the recent success of neural network methods in various domains, we further design a hierarchical attention framework (see Figure~\ref{fig-haf}) based on graph neural network models, which integrates these two stages and can be trained in an end-to-end manner. In Section~\ref{sec-exp-rmse}, we can see that HAF outperforms FMG by a large margin, which demonstrates the power of neural network models.


\begin{remark}
The main contribution of this paper is the novel and systematical paradigm to fuse different types of side information over HIN for the popular CF based recommendation problems. This paradigm includes two types of frameworks: one is the ``MF + FM'' framework in Figure \ref{fig-framework}, and the other is the neural framework in Figure \ref{fig-haf}. Despite the effectiveness of these two frameworks, another problem is facing all HIN-based recommendation methods in practice, which is that \textit{how to construct a HIN in a recommendation scenario.} In the context of the rating prediction task in CF based recommendation, we refer to \textit{any potentially useful auxiliary information} beyond the rating matrix as side information, e.g., the social connections among users, meta-data of items, and reviews including texts and images. In general, those in discrete values, e.g., category, city, state in Figure \ref{fig-intro-example}, can be directly modeled as nodes in a HIN, however,  there are some types of side information which cannot be easily to modeled in a HIN. For example, to make use of the textual content of reviews in Yelp (Figure \ref{fig-intro-example}), we extract keywords from each review as \textit{Aspect} node in a HIN by natural language processing techniques (see the network schema in Figure \ref{fig-hin-schema}), however, it remains to be a challenging and unknown problem to incorporate images into the constructed HIN. This problem can be a potential future direction beyond this work, not only in recommendation scenarios, but also in other HIN-based problems, e.g., intent recommendation~\cite{fan2019metapath}, fraud detection~\cite{hu2019cash}, malware detection in software systems~\cite{hou2017hindroid,fan2018gotcha}. Once a HIN is constructed, the proposed methods in this paper can naturally be adopted, which demonstrates the practical values in broader domains. 



\end{remark}

\section{Metagraph Construction and Feature Extraction}
\label{sec-mf-part}

In this section, we elaborate on the MF part for metagraph based feature extraction. In Section~\ref{sec-mg-construction}, we introduce the method for constructing metagraphs in HIN. Then, we show how to compute the user-item similarity matrices in Section~\ref{sec-mg-sim}. Finally, in Section~\ref{sec-mg-latent-features}, we obtain latent features from the user-item matrices using MF-based approaches. The main novelty of our approach is the design of the MF part, which extracts and combines latent features from each metagraph before they are fed to the FM part. Further, as existing methods are only for computing similarity matrices based on metapath, we show how similarity can be computed for metagraph.

\subsection{Construction of Metagraphs}
\label{sec-mg-construction}
We first give the definitions of HIN, network schema for HIN, and Metagraph~\cite{sun2011pathsim,Huang2016MetaSC,Fang2016SemanticPS,fang2019metagraph}.
Then, we introduce how to compute metagraph based similarities between users and items in a HIN.

\begin{definition}[Heterogeneous Information Network]
A \textbf{heterogeneous information network} (HIN) is a graph ${\mathcal G} = ({\mathcal V}, {\mathcal E})$ with an entity type mapping $\phi$: ${\mathcal V} \to \mathcal A$ and a relation type mapping $\psi$: ${\mathcal E} \to \mathcal R$, where ${\mathcal V}$ denotes the entity set, ${\mathcal E}$ denotes the link set, $\mathcal A$ denotes the entity type set, and $\mathcal R$ denotes the relation type set, and the number of entity types $|\mathcal A|>1$ or the number of relation types $|\mathcal R|>1$.
\end{definition}

\begin{definition}[Network schema]

Given a HIN ${\mathcal G} = ({\mathcal V}, {\mathcal E})$ with the entity type mapping $\phi$: ${\mathcal V} \to \mathcal A$ and the relation type mapping $\psi$: $\mathcal E \to \mathcal R$, the \textbf{network schema} for network $\mathcal{G}$, denoted by $\mathcal T_{\mathcal G} = (\mathcal A, \mathcal R)$, is a graph, in which nodes are entity types from $\mathcal A$ and edges are relation types from $\mathcal R$.
\end{definition}

Figures~\ref{fig-intro-example} and \ref{fig-hin-schema} show an example of HIN and its network schema from the Yelp dataset respectively. 
We can see that we have different types of nodes, e.g., \textit{User, Review, Restaurant}, and different types of relations, e.g., \textit{Write} and \textit{CheckIn}. 
The network schema defines the relations between node types, e.g., \textit{User} \textit{CheckIn} \textit{Restaurant}, and \textit{Restaurant} \textit{LocateIn} \textit{City}. 
Thus, we can see that HIN is a flexible way for representing various information in a unified manner. The definition of metagraph is given below.

\begin{definition}[Metagraph]
A metagraph ${\mathcal M}$ is a directed acyclic graph (DAG) with a single source node $n_s$ (i.e., with in-degree 0) and a single sink (target) node $n_t$ (i.e., with out-degree 0), defined in a HIN  ${\mathcal G} = ({\mathcal V},{\mathcal E})$. Formally, $\mathcal{M} = (\mathcal{V}_M, \mathcal{E}_M, \mathcal{A}_M, \mathcal{R}_M, n_s, n_t)$, where $\mathcal{V}_M \subseteq \mathcal{V}$ and $\mathcal{E}_M \subseteq \mathcal{E} $ are constrained by $\mathcal{A}_M \subseteq \mathcal{A}$ and $\mathcal{R}_M \subseteq \mathcal{R}$, respectively. 
\end{definition}

\begin{figure}[t]
	\centering
	\includegraphics[width=0.7\textwidth]{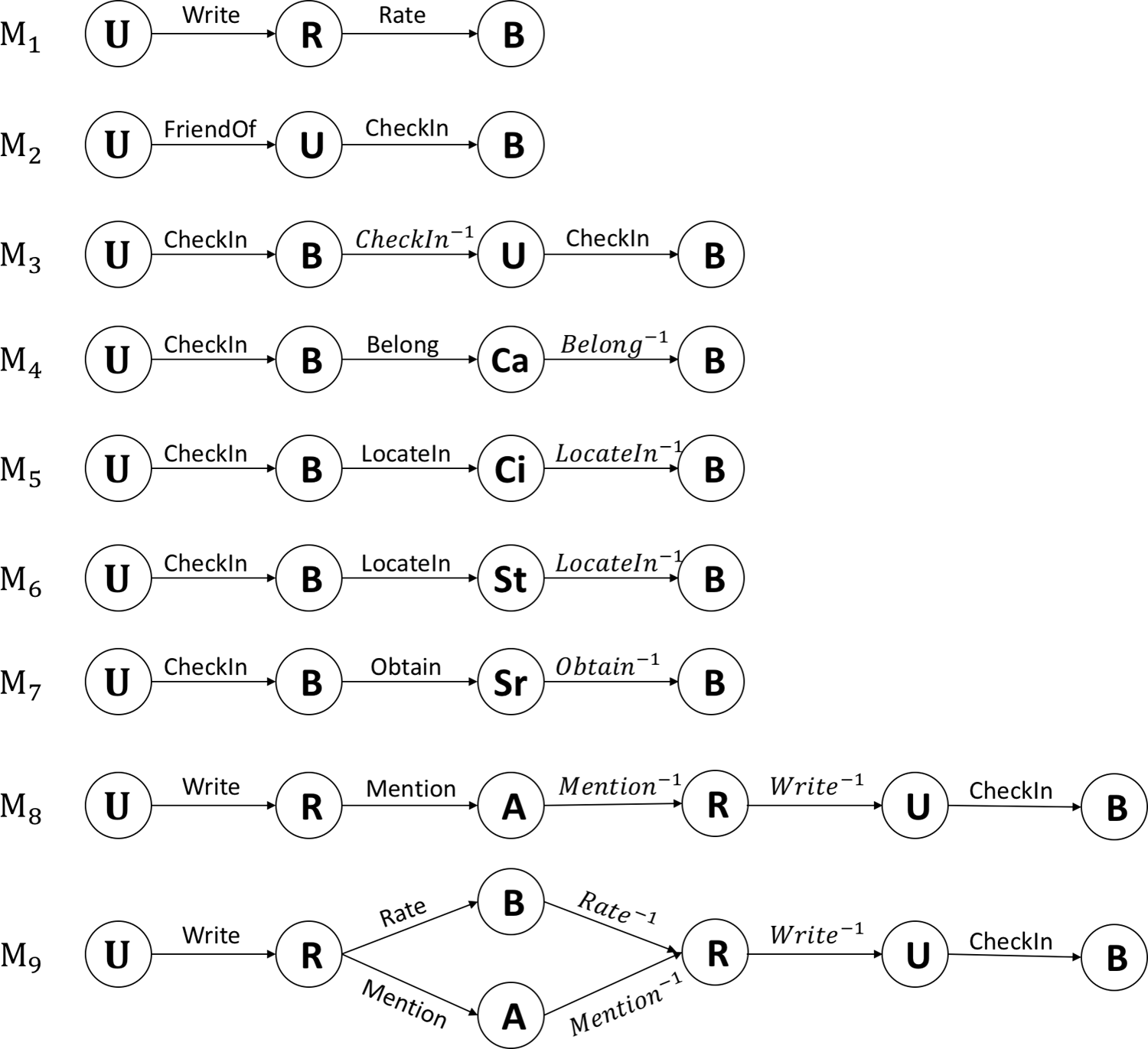}
	\vspace{-5px}
	\caption{Metagraphs used for the Yelp dataset (\textbf{Ca}: Category; \textbf{Ci}: City; \textbf{St}: State; \textbf{Sr}: Star, the average number of stars a business obtained).}
	\label{fig-yelp-metagraph}
\end{figure}

\begin{figure}[t]
	\centering
	\includegraphics[width=0.7\textwidth]{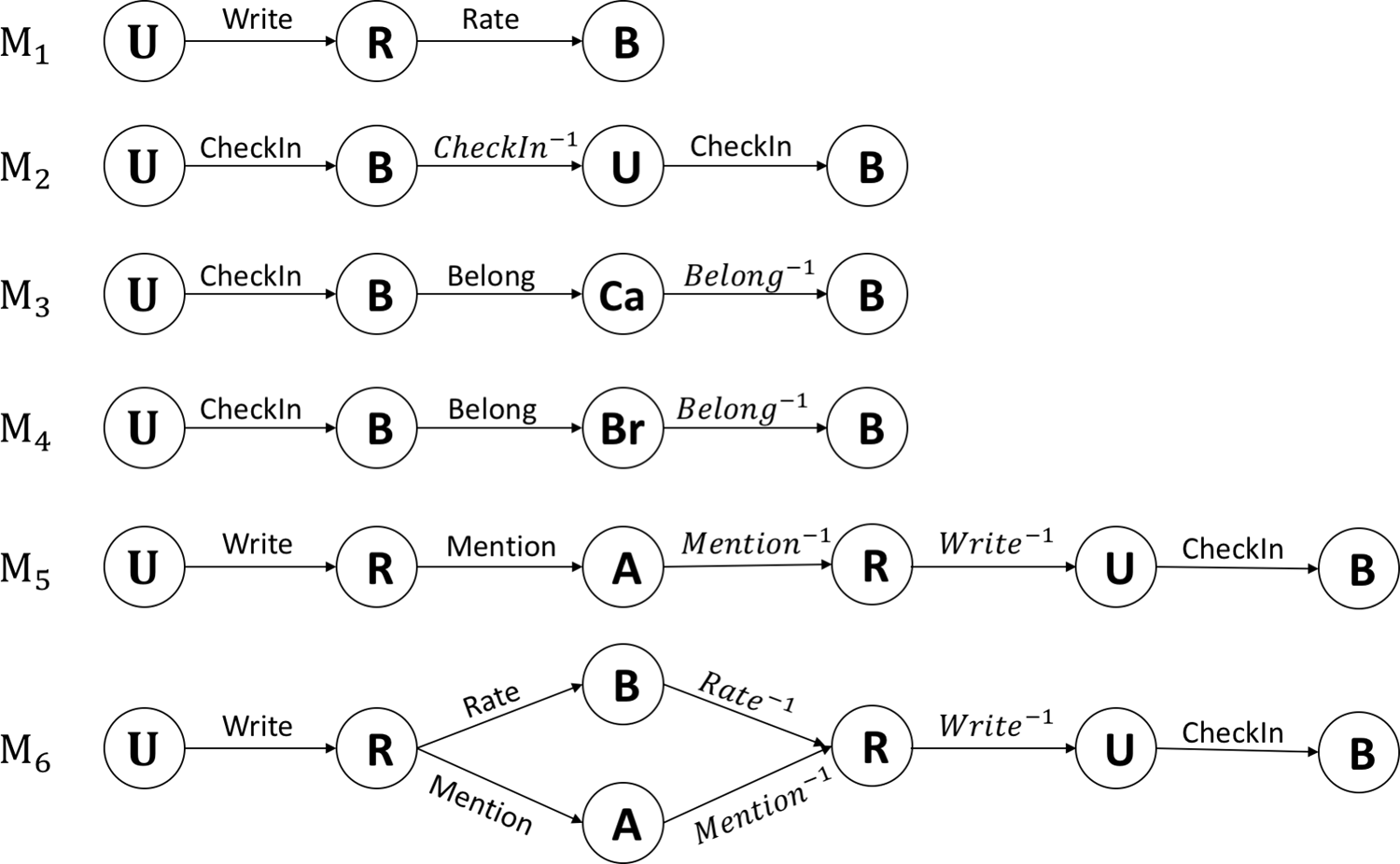}
	\vspace{-5px}
	\caption{Metagraphs used for the Amazon dataset (Ca: Category; Br: Brand of the item).}
	\label{fig-amazon-metagraph}
\end{figure}

As in~\cite{Fang2016SemanticPS,Huang2016MetaSC,fang2019metagraph}, 
compared to metapath, metagraph can capture more complex semantics underlying the similarities between users and items. 
In fact, metapath is a special case of metagraph. Thus, in this paper, we introduce the concept of metagraph for HIN-based RS. In Figures~\ref{fig-yelp-metagraph} and \ref{fig-amazon-metagraph}, we show the metagraphs on Yelp and Amazon datasets, respectively, used in our experiments. In these figures, $\mathcal{R}^{-1}$ represents the reverse relation of $\mathcal{R}$. For example, for $M_3$ in Figure~\ref{fig-yelp-metagraph}, $B \xrightarrow{CheckIn^{-1}} U$ means $U$ checks in a business $B$. From Figure~\ref{fig-yelp-metagraph} and \ref{fig-amazon-metagraph}, we can see that each metagraph has only one source ($U$) and one target ($B$) node, representing a user and an item in the recommendation scenario.

\subsubsection{Practical Suggestions}
\label{sec:sugg}

Since there could be many metagraphs in a HIN and they are not equally effective, we give three guidelines for the selection of metagraphs:
1) All metagraphs designed are from the network schema.
2) Domain knowledge is helpful because some metagraphs correspond to traditional recommendation strategies that have been proven to be effective~\cite{yu2014personalized,shi2015semantic}.
For example, $\mathcal{M}_2$ and $\mathcal{M}_3$ in Figure~\ref{fig-yelp-metagraph}, respectively, represent social recommendation and the well-known user-based CF. In practice, an understanding of existing recommendation strategies and application semantics is essential for the design of good metagraphs; 
3) It is better to construct shorter metagraphs. In~\cite{sun2011pathsim}, the authors have shown that longer metapaths decrease the performance because they tend to have more noises. This result is applicable to metagraphs as well.

\subsection{Computation of Similarity Matrices}
\label{sec-mg-sim}
We use ${\mathcal M}_3$ and ${\mathcal M}_9$ in Figure~\ref{fig-yelp-metagraph} to illustrate the computation of metagraph based similarities.
In previous works, 
commuting matrices~\cite{sun2011pathsim} have been employed to compute the count-based similarity matrix of a metapath.
Suppose we have a metapath $\mathcal{P} = (\mathbb{A}_1,\mathbb{A}_2,\ldots,\mathbb{A}_l)$, 
where $\mathbb{A}_i$'s are node types in $\mathcal{A}$ and denote the adjacency matrix between type $\mathbb{A}_i$ and type $\mathbb{A}_j$ by $\tW_{\mathbb{A}_i\mathbb{A}_j}$.
Then the commuting matrix for $\mathcal{P}$ is defined by the multiplication of a sequence of adjacency matrices:
\begin{align*}
\tC_P =\tW_{\mathbb{A}_1,\mathbb{A}_2} \tW_{\mathbb{A}_2,\mathbb{A}_3} \cdots \tW_{\mathbb{A}_{l-1}, \mathbb{A}_l},
\end{align*}
where $\tC_P(i,j)$, the entry in the $i$-th row and $j$-th column, represents the number of path instances between object $x_i \in \mathbb{A}_1$ and object $x_j \in \mathbb{A}_l$ under $\mathcal{P}$. For example, 
for ${\mathcal M}_3$ in Figure~\ref{fig-yelp-metagraph}, $\tC_{M_3} = \tW_{UB}  \tW_{UB}^{\top}  \tW_{UB}$, 
where $\tW_{UB}$ is the adjacency matrix between type $U$ and type $B$, and $\tC_{M_3}(i,j)$ represents the number of instances of $\mathcal{M}_3$ between user $u_i$ and item $b_j$.
In this paper, for a metagraph $\mathcal{M}$,
the similarity between a source object and a target object is defined as
the number of instances of $\mathcal{M}$ connecting the source and target objects.  In the remainder of this paper, we adopt the term similarity matrix instead of commuting matrix for clarity.

From the above introduction, we can see that metapath based similarity matrix is easy to compute. However, for metagraphs, the problem is more complicated. For example, consider ${\mathcal M}_9$ in Figure~\ref{fig-yelp-metagraph}, there are two ways to pass through the metagraph, which are $(U,R,A,R,U,B)$ and $(U,R,B,R,U,B)$.
Note that $R$ represents the entity type {\it Review} in HIN. In the path $(U,R,A,R,U,B)$, $(R,A,R)$ means if two reviews mention the same $A$ ({\it Aspect}), then they have some similarity.
Similarly, in $(U,R,B,R,U,B)$, $(R,B,R)$ means if two reviews rate the same $B$ ({\it Business}), they have some similarity too.
We should decide how similarity should be defined when there are multiple ways to pass through the metagraph from the source node to the target node.
We can require a flow to pass through either path or both paths in order to be considered in similarity computation.
The former strategy simply splits a metagraph into multiple metapaths, thus suffering from information loss. Therefore, we adopt the latter, but it requires one more matrix operation in addition to simple multiplication, 
i.e, element-wise product.
Algorithm~\ref{alg-metagraph-commu-mat} depicts the algorithm for computing count-based similarity based on ${\mathcal M}_9$ in Figure~\ref{fig-yelp-metagraph}.
After obtaining $\tC_{S_r}$, we can get the whole similarity matrix $\tC_{M_9}$ by multiplying the sequence of matrices along $\tC_{M_9}$. In practice, not limited to ${\mathcal M}_9$ in Figure~\ref{fig-yelp-metagraph}, the metagraph defined in this paper can be computed by two operations (Hadamard product and multiplication) on the corresponding matrices.

\begin{algorithm}[ht]
\caption{Computing similarity matrix based on $\mathcal{M}_9$.}
\small
\begin{algorithmic}[1]
	\STATE{Compute $\tC_{P_1}: \tC_{P_1} = \tW_{RB}  \tW_{RB}^{\top}$;}
	\STATE{Compute $\tC_{P_2}: \tC_{P_2} = \tW_{RA}  \tW_{RA}^{\top}$;}
	\STATE{Compute $\tC_{S_r}: \tC_{S_r} = \tC_{P_1} \odot \tC_{P_2}$;}
	\STATE{Compute $\tC_{M_9}: \tC_{M_9} = \tW_{UR}  \tC_{S_r}  \tW_{UR}^{\top}  \tW_{UB}$.}	
\end{algorithmic}
\label{alg-metagraph-commu-mat}
\end{algorithm}

By computing the similarities between all users and items for the $l$-th metagraph $\mathcal{M}$, 
we can obtain a user-item similarity matrix
$\textbf{R}^l \in \mathbb{R}^{m \times n}$,
where $\textbf{R}^l_{ij}$ represents the similarity between user $u_i$ and item $b_j$ along the metagraph, 
and $m$ and $n$ are the number of users and items, respectively. 
Note that $\textbf{R}^l_{ij} = \tC_{M_l}(i,j)$
\footnote{To maintain consistency with the remaining sections, we change the notation $\tC$ into $\bR$.} 
if $\tC_{M_l}(i,j) > 0$ and 0 otherwise. By designing $L$ metagraphs, we can get $L$ different user-item similarity matrices, denoted by $\textbf{R}^1, \ldots, \textbf{R}^L$.

\subsection{Latent Feature Generation}
\label{sec-mg-latent-features}

In this part, we elaborate on how to generate latent features for users and items from the $L$ user-item similarity matrices. Since the similarity matrices are usually very sparse, using the matrices directly as features will lead to the high-dimensional learning problem, resulting in overfitting.
Motivated by the success of low-rank matrix completion
\cite{mnih2007probabilistic,koren2008factorization,candes2009exact},
we propose to generate latent features using these methods.

Specifically, the nonzero elements in a similarity matrix are treated as observations and the others are taken as missing values.
Then we find a low-rank approximation to this matrix.
Matrix factorization (MF) \cite{koren2008factorization,mnih2007probabilistic} 
and nuclear norm regularization (NNR) \cite{candes2009exact} are two popular approaches
for matrix completion.
Generally,
MF leads to nonconvex optimization problems,
while NNR leads to convex optimization problems.
NNR is easier to optimize and has better theoretical guarantee on the recovery performance than MF.
Empirically, NNR usually has better performance and the recovered rank is often much higher than that of MF~\cite{yao2015accelerated}. In this paper, we generate metagraph based latent features with both methods and conduct experiments to compare their performance (shown in Section~\ref{sec-exp-mf-vs-nn}). The technical details of these two methods are introduced in the remaining part of this section.

\subsubsection{Matrix Factorization}

Consider a user-item similarity matrix $\mathbf{R} \in \mathcal{R}^{m \times n}$,
let the observed positions be indicated by $1$'s in $\mathbf{\Omega} \in \{0, 1\}^{m \times n}$, 
i.e., $\left[ \SO{\textbf{X}} \right]_{ij} = \textbf{X}_{ij}$ if $\mathbf{\Omega}_{ij} = 1$ and 0 otherwise. 
$\mathbf{R}$ is factorized as a product of $\mathbf{U} \in \mathcal{R}^{m \times F}$ 
and $\mathbf{B} \in \mathcal{R}^{n \times F}$ by solving the following optimization problem:
\begin{equation}
\label{eq-mf}
\min_{\textbf{U},\textbf{B}}\frac{1}{2} \NM{ \SO{\textbf{U}\textbf{B}^{\top} - \textbf{R}} }{F}^2 + 
\frac{\mu}{2}\left( \NM{\textbf{U}}{F}^2 + \NM{ \textbf{B} }{F}^2\right) ,
\end{equation}
where $F \ll \min\left( m, n \right)$ is the desired rank of $\mathbf{R}$,
and $\mu$ is the hyper-parameter controlling regularization.

We adopt the gradient descent based approach for optimizing \eqref{eq-mf},
which is popular in RSs~\cite{koren2008factorization,mnih2007probabilistic}.
After optimization,
we take $\textbf{U}$ and $\textbf{B}$ as the latent features of users and items, respectively. 

\subsubsection{Nuclear Norm Regularization}
\label{sec:nnr}

Although MF can be simple, \eqref{eq-mf} is not a convex optimization problem, so there is no rigorous guarantee on the recovery performance.
This motivates our adoption of nuclear norm, 
which is defined as the sum of all singular values of a matrix. 
It is also the tightest convex envelope to the rank function.
This leads to the following nuclear norm regularization (NNR) problem:
\begin{equation}
\min_{\mathbf{X}}
\frac{1}{2} \NM{P_\mathbf{\Omega}( \mathbf{X} - \textbf{R})}{F}^2 + \mu \NM{\mathbf{X}}{*},
\label{eq-nnr}
\end{equation}
where $\mathbf{X}$ is the low-rank matrix to be recovered, and $\mu$ is the hyper-parameter controlling regularization.
Nice theoretical guarantee has been developed for \eqref{eq-nnr},
which shows that $\mathbf{X}$ can be exactly recovered given sufficient observations \cite{candes2009exact}.
These advantages make NNR popular for low-rank matrix approximation \cite{candes2009exact}. Thus, we adopt \eqref{eq-nnr} to generate latent features,
using the state-of-the-art AIS-Impute algorithm \cite{yao2015accelerated} in optimizing \eqref{eq-nnr}.
It has fast $O(1/T^2)$ convergence rate,
where $T$ is the number of iterations,
with low per-iteration time complexity.
In the iterations,
a Singular Value Decomposition (SVD) of $\mathbf{X} = \mathbf{P} \mathbf{\Sigma} \mathbf{Q}^{\top}$ is maintained
($\mathbf{\Sigma}$ only contains the nonzero singular values).
When the algorithm stops,
we take $\mathbf{U} = \mathbf{P} \mathbf{\Sigma}^{\frac{1}{2}}$
and $\mathbf{B} = \mathbf{Q} \mathbf{\Sigma}^{\frac{1}{2}}$ as user and item latent features, respectively.

\subsection{Complexity Analysis}
We analyze the time complexity of the MF part,
which includes similarity matrix computation and latent feature generation. 
For similarity matrix computation, the core part is matrix multiplication. 
Since the adjacency matrices tend to be very sparse, 
they can be implemented very efficiently as sparse matrices.
Moreover, for MF and NNR, according to~\cite{yao2015accelerated,yao2018large}, the computation cost in each iteration is $O(|| \Omega ||_1 F + (m+n)F)$ and $O(|| \Omega ||_1 F + (m+n)F^2)$, respectively, where $|| \Omega ||_1$ is the number of nonzero elements in the similarity matrix, $m$ and $n$ are the dimensions of the similarity matrix, and $F$ is the rank used in the factorization of the similarity matrix.

\section{Metagraph-based Features Fusion and Selection}
\label{sec-framework-model}
In this section, we describe the FM part for fusing multiple groups of metagraph based latent features. 
Existing HIN-based RS methods~\cite{yu2014personalized,shi2015semantic} only use a linear combination of different metapath based features and thus ignore the interactions among features. 
To resolve this limitation, we apply FM to capture the interactions among metagraph based latent features and non-linear interactions among features (i.e., second-order interactions) when fusing various side information in HIN. In Section~\ref{sec-fm-pred}, we show how FM performs prediction utilizing the metagraph based latent features.
Then we introduce two regularization terms in Section~\ref{sec-framework-convex}, which can achieve automatic metagraph selection. In Section~\ref{sec:opt}, we depict the objective function and propose two optimization methods for it.

\subsection{Combining Latent Features with FM}
\label{sec-fm-pred}

In this section, we introduce our FM-based algorithm for fusing different groups of latent features. As described in Section~\ref{sec-mg-latent-features}, we obtain $L$ groups of latent features of users and items, denoted by $\textbf{U}^{1}$, $\textbf{B}^{1}$, $\dots$, $\textbf{U}^{L}$, $\textbf{B}^{L}$, from $L$ metagraph based user-item similarity matrices.
For a sample $\mathbf{x}^n$ in the observed ratings, i.e., a pair of user and item, denoted by $\textbf{u}_i$ and $\textbf{b}_j$, respectively, we concatenate all of the corresponding user and item features from the $L$ metagraphs:
\begin{align}
\mathbf{x}^n = 
\big[ \underbrace{\textbf{u}^{1}_i, \cdots, \textbf{u}^{L}_i}_{ \sum_{l=1}^{L}F_l }
,
\underbrace{\textbf{b}^{1}_j, \cdots, \textbf{b}^{L}_j}_{ \sum_{l=1}^{L}F_l } \big]
\in \mathbb{R}^{d},
\label{eq-fm-graph}
\end{align}
where $d = 2\sum_{l=1}^{L}F_l$, and $F_l$ is the rank of the factorization of the similarity matrix for the $l$-th metagraph obtained with \eqref{eq-mf} or \eqref{eq-nnr}. $\textbf{u}^{l}_i$ and $\textbf{b}^{l}_j$, respectively, represent user and item latent features generated from the $l$-th metagraph, and $\mathbf{x}^n$ is a $d$-dimension vector representing the feature vector of the $n$-th sample after concatenation.

Given all of the features in (\ref{eq-fm-graph}), the predicted rating for the sample $\mathbf{x}^n$ based on FM~\cite{rendle2012fm} is computed as follows:
\begin{equation}
\hat{y}^n (\mathbf{w}, \mathbf{V}) 
= b + \sum\nolimits_{i=1}^{d}w_ix^n_i + \sum\nolimits_{i=1}^{d}\sum\nolimits_{j=i+1}^{d}\langle\mathbf{v}_i,\mathbf{v}_j\rangle x^n_ix^n_j,
\label{eq-rating-fm}
\end{equation}
where $b$ is the global bias, and $\mathbf{w} \in \mathbb{R}^{d}$ represents the first-order weights of the features. $\mathbf{V} = [\mathbf{v}_i] \in \mathbb{R}^{d \times K}$ represents the second-order weights for modeling the interactions among the features, 
and $\mathbf{v}_i$ is the $i$-th row of the matrix $\mathbf{V}$, which describes the $i$-th variable with $K$ factors. $\mathbf{x}^n_i$ is the $i$-th feature in $\mathbf{x}^n$. The parameters can be learned by minimizing the mean square loss:
\begin{equation}
\ell\left( \mathbf{w}, \mathbf{V} \right) =
\frac{1}{N}
\sum\nolimits_{n = 1}^N \left( y^n - \hat{y}^n(\mathbf{w}, \mathbf{V}) \right)^2,
\label{eq:loss}
\end{equation}
where $y^n$ is an observed rating for the $n$-th sample, and $N$ is the number of all observed ratings.

\subsection{Metagraph Selection with Group Lasso}
\label{sec-framework-convex}
We need to tackle two problems when FM is applied to metagraph based latent features. The first problem is that noise may arise when there are too many metagraphs, thus impairing the predicting capability of FM. This is because not all metagraphs are useful for recommendation because the semantics captured in a metagraph may have little effect on recommendation behavior in the real world.
The second problem is computational cost. All of the features are generated by MF, which means that the design matrix (i.e., features fed to FM) is dense. It increases the computational cost for learning the parameters of the model and that of online recommendation. To alleviate these two problems, we propose two novel regularization terms to automatically select useful metagraphs during training process. They can be categorized into convex and nonconvex regularizations, and either of them enables our model to automatically select useful metagraphs during the training process.

\subsubsection{Convex Regularization}
The convex regularizer is the $\ell_{2,1}$-norm regularization,  i.e., group lasso regularization~\cite{jacob2009group}, which is a feature selection method on a group of variables. 
Given the pre-defined non-overlapping $G$ groups $\left\lbrace \mathcal{I}_1, \dots, \mathcal{I}_G \right\rbrace $ on the parameter $\mathbf{p}$,
the regularization is defined as follows.
\begin{equation}
\phi(\mathbf{p}) = \sum\nolimits_{g=1}^{G} \eta_g \NM{\mathbf{p}_{\mathcal{I}_g}}{2},
\label{eq-l21-norm}
\end{equation}
where $\NM{\cdot}{2}$ is the $\ell_2$-norm, and $\eta_g$ is a hyper-parameter. 
In our model, the groups correspond to the metagraph based features. For example, $\mathbf{U}^{l}$ and $\mathbf{B}^{l}$ are the user and item latent features generated by the $l$-th metagraph. For a pair of user $i$ and item $j$, the latent features are $\mathbf{u}^{l}_i$ and $\mathbf{b}^{l}_j$.
There are two corresponding groups of variables in $\mathbf{w}$ and $\mathbf{V}$ according to (\ref{eq-rating-fm}). Thus,
with $L$ metagraphs, $\textbf{w}$ and $\textbf{V}$ each has $2L$ groups of variables.

For the first-order parameters $\mathbf{w}$ in (\ref{eq-rating-fm}), which is a vector, group lasso is applied to the subset of variables in $\mathbf{w}$. Then we have:
\begin{equation}
\hat{\phi}(\mathbf{w}) = \sum\nolimits_{l = 1}^{2L} \hat{\eta}_l \NM{ \mathbf{w}^l }{2},
\label{eq-glasso-w}
\end{equation}
where $\mathbf{w}^l \in \mathbb{R}^{F_l}$, which models the weights for a group of user or item features from one metagraph, and $\hat{\eta}_l$ is a hyper-parameter.
For the second-order parameters $\mathbf{V}$ in (\ref{eq-rating-fm}), we have the regularizer as follows:
\begin{equation}
\bar{\phi}(\mathbf{V}) = \sum\nolimits_{l = 1}^{2L} \bar{\eta}_l \NM{ \mathbf{V}^l }{F},
\label{eq-glasso-v}
\end{equation} 
where  $\mathbf{V}^l  \in \mathbb{R}^{F_l \times K}$, the $l$-th block of $\textbf{V}$ corresponds to the $l$-th metagraph based features in a sample, and $\NM{\cdot}{F}$ is the Frobenius norm.

\subsubsection{Nonconvex Regularization}
\label{sec:ncvxreg}

While convex regularizers usually make optimization easy, they often lead to biased estimation. 
For example, 
in sparse coding,  
the solution obtained by the $\ell_1$-regularizer is often not as sparse and accurate compared to capped-$\ell_1$ penalty~\cite{zhang2010analysis}.
Besides, in low-rank matrix learning, the estimated rank obtained with the nuclear norm regularizer is often very high \cite{yao2018large}. To alleviate these problems, a number of nonconvex regularizers, which are variants of the convex $\ell_1$-norm, have been recently proposed~\cite{yao2016efficient,yao2018large}.
Empirically, these nonconvex regularizers usually outperform the convex ones.
Motivated by the above observations,
we propose to use nonconvex variant of \eqref{eq-glasso-w} and \eqref{eq-glasso-v} as follows:
\begin{equation}
\hat{\psi}
(\mathbf{w}) 
\! = \! \sum\nolimits_{l = 1}^{2L} \hat{\eta}_l \kappa \left( \NM{ \mathbf{w}^l }{2} \right),
\quad
\bar{\psi}
(\mathbf{V}) 
\! = \! \sum\nolimits_{l = 1}^{2L} \bar{\eta}_l \kappa \left( \NM{ \mathbf{V}^l }{F} \right),
\label{eq:gpreg}
\end{equation} 
where $\kappa$ is a nonconvex penalty function.
We choose $\kappa\left( |\alpha| \right) = \log\left( 1 + |\alpha| \right)$ as the log-sum-penalty (LSP)~\cite{candes2008enhancing},
since it has been shown to give the best empirical performance on learning sparse vectors \cite{yao2016efficient} and low-rank matrices 
\cite{yao2018large}.

\subsubsection{Comparison with Existing Methods}
\label{sec-comparsion}

\citet{yu2014personalized} studied recommendation techniques based in HINs and applied matrix factorization to generate latent features from metapaths. Ratings are predicted using a weighted ensemble of the dot products of user and item latent features from every single metapath:
$\hat{r}(\textbf{u}_i,\textbf{b}_j) = \sum_{l=1}^{L} \theta_l \cdot \textbf{u}^{l}_i ( \textbf{b}^{l}_j )^{\top}$, where $\hat{r}(\textbf{u}_i,\textbf{b}_j)$ is the predicted rating for user $u_i$ and item $b_j$ and $\textbf{u}^{l}_i$ and $\textbf{b}^{l}_j$ are the latent features for $u_i$ and item $b_j$ from the $l$-th metapath, respectively. $L$ is the number of metapaths used, and $\theta_l$ is the weight for the $l$-th metapath latent features.  However, the predicting method is not adequate, as it fails to capture the interactions between features across different metapaths, and between features within the same metapath, resulting in a decrease of the prediction performance for all of the features. In addition, previous works on FM \cite{rendle2012fm,hong2013co,yan2014coupled} 
only focus on the selection of one row or column of the second-order weight matrix,
while $\bar{\phi}$ in our method selects a block of rows or columns (defined by metagraphs).
Moreover, we are the first to adopt nonconvex regularization, i.e., $\bar{\psi}$, 
for weight selection in FM.

\subsection{Model Optimization} 
\label{sec:opt}

Combining \eqref{eq:loss} and \eqref{eq:gpreg},
we define our FM with Group lasso (FMG) model with the following objective function:
\begin{align}
h(\mathbf{w}, \mathbf{V})
\! = \!
\frac{1}{N} \sum\nolimits_{n = 1}^N (y^n \! - \! \hat{y}^n(\mathbf{w}, \mathbf{V}))^2 
\! + \! \hat{\lambda} \hat{\psi}(\mathbf{w}) 
\! + \! \bar{\lambda} \bar{\psi}(\mathbf{V}).
\label{eq-obj}
\end{align}
Note that when $\kappa(\alpha) = |\alpha|$ in \eqref{eq:gpreg},
we get back \eqref{eq-glasso-w} and \eqref{eq-glasso-v}.
Thus, we directly use the nonconvex regularization in \eqref{eq-obj}.

We can see that $h$ is nonsmooth due to the use of $\hat{\phi}^{\textbf{w}}$ and $\hat{\phi}^{\textbf{V}}$,
and nonconvex due to the nonconvexity of loss $\ell$  on $\mathbf{w}$ and $\mathbf{V}$.
To alleviate the difficulty on optimization,
inspired by \cite{yao2016efficient},
we propose to reformulate \eqref{eq-obj} as follows:
\begin{align}
\bar{h}(\mathbf{w}, \mathbf{V})
=
\bar{\ell}\left( \mathbf{w}, \mathbf{V} \right) 
+ \kappa_0 \hat{\lambda} \hat{\phi}(\mathbf{w}) 
+ \kappa_0 \bar{\lambda} \bar{\phi}(\mathbf{V}),
\label{eq:equiv}
\end{align}
where $\bar{\ell}\left( \mathbf{w}, \mathbf{V} \right) = \ell( \mathbf{w}, \mathbf{V} ) + g( \mathbf{w}, \mathbf{V} )$,
$\kappa_0 = \lim_{\beta \rightarrow 0^+} \kappa'(|\beta|)$
and
\begin{align*}
g( \mathbf{w}, \mathbf{V} )
= \hat{\lambda} \left[ \hat{\psi}(\mathbf{w}) - \kappa_0 \hat{\phi}(\mathbf{w}) \right] 
+ \bar{\lambda} \left[ \bar{\psi}(\mathbf{V}) - \kappa_0 \bar{\phi}(\mathbf{V}) \right].
\end{align*}
Note that $\bar{h}$ is equivalent to $h$ based on Proposition 2.1 in \cite{yao2016efficient}.
A very important property for the augmented loss $\bar{\ell}$ is that it is still smooth.
As a result,
while we are still optimizing a nonconvex regularized problem,
we only need to deal with convex regularizers.

In Section~\ref{sec:nmAPG}, we show how the reformulated problem can be solved by the state-of-the-art proximal gradient algorithm \cite{li2015accelerated};
moreover,
such transformation enables us to design a more efficient optimization algorithm with convergence guarantee based on variance reduced methods \cite{xiao2014proximal}.
Finally,
the time complexity of the proposed algorithms is analyzed in Section~\ref{sec:timecomp}.

\begin{remark}
nmAPG was previously used in our paper \cite{zhao2017meta}. Here, we show that it can still be applied to the new model \eqref{eq:equiv}. Besides, we further propose to use SVRG and show in Section~\ref{sec-exp-svrg-vs-nmAPG} that it is much more efficient than nmAPG.
\end{remark}

\subsubsection{Using nmAPG Algorithm}
\label{sec:nmAPG}

To tackle the nonconvex nonsmooth objective function \eqref{eq:equiv}, we propose to adopt the PG algorithm~\cite{parikh2014proximal} and, specifically,
the state-of-the-art non-monotonic accelerated proximal gradient (nmAPG) algorithm \cite{li2015accelerated}. It targets at optimization problems of the form:
\begin{align}
\min_\mathbf{x} F(\mathbf{x}) \equiv f(\mathbf{x}) + g(\mathbf{x}),
\label{eq:compopt}
\end{align}
where $f$ is a smooth (possibly nonconvex) loss function and 
$g$ is a regularizer (can be nonsmooth and nonconvex).
To guarantee the convergence of nmAPG,
we also need $\lim_{\NM{\mathbf{x}}{2} \rightarrow \infty} F(\mathbf{x}) = \infty$, 
$\inf_{\mathbf{x}} F(\mathbf{x}) > - \infty$, and there exists at least one solution to the proximal step, i.e., 
$\Px{\gamma g}{\mathbf{z}} = \arg\min_{\mathbf{x}} \nicefrac{1}{2} \| \mathbf{x} - \mathbf{z} \|_2^2 + \gamma g(\mathbf{x})$,
where $\gamma \ge 0$ is a scalar \cite{li2015accelerated}.

The motivation of nmAPG is two fold.
First, nonsmoothness comes from the proposed regularizers, which can be efficiently handled if the corresponding proximal steps have cheap closed-form solution.
Second, the acceleration technique is useful for significantly speeding up first order optimization algorithms \cite{yao2016efficient,li2015accelerated,yao2017niapg}, 
and nmAPG is the state-of-the-art algorithm which can deal with general nonconvex problems with sound convergence guarantee.
The whole procedure is given in Algorithm~\ref{alg-nmAPG}.
Note that while both $\hat{\phi}$ and $\bar{\phi}$ are nonsmooth in \eqref{eq:equiv},
they are imposed on $\mathbf{w}$ and $\mathbf{V}$ separately.
Thus, 
for any $\alpha, \beta \ge 0$,
we can also compute proximal operators independently for these two regularizers following~\cite{parikh2014proximal}:
\begin{align}
\Px{\alpha \hat{\phi} + \beta \bar{\phi}}{\mathbf{w}, \mathbf{V}}
= 
\big( 
\Px{\alpha \hat{\phi}}{\mathbf{w}},
\Px{\beta \bar{\phi}}{\mathbf{V}}
\big).
\label{eq:proxsep}
\end{align}
These are performed in steps~5 and 10 in Algorithm~\ref{alg-nmAPG}. 
The closed-form solution of the proximal operators can be obtained easily from Lemma~\ref{lem:solution} below.
Thus, each proximal operator can be solved in one pass of all groups.

\begin{lemma}[\cite{parikh2014proximal}] \label{lem:solution}
	The closed-form solution of $\mathbf{p}^* = \Px{\lambda \phi}{\mathbf{z}}$ 
	($\phi$ is defined in \eqref{eq-l21-norm}) is given by
	$\mathbf{p}^*_{\mathcal{I}_g} 
	= \max\left( 1 - \nicefrac{\eta_g }{ \NM{\mathbf{z}_{\mathcal{I}_g}}{2} }, 0 \right) \mathbf{z}_{\mathcal{I}_g}$
	for all $g = 1, \dots, G$.
\end{lemma}

It is easy to verify that the above assumptions are satisfied by our objective $h$ here.
Thus, Algorithm~\ref{alg-nmAPG} is guaranteed to produce a critical point for \eqref{eq:equiv}.

\begin{algorithm}[ht]
	\caption{nmAPG algorithm for \eqref{eq:equiv}.} 
	\small
	\begin{algorithmic}[1]
		\STATE{Initiate $\textbf{w}_0, \textbf{V}_0$ as Gaussian random matrices;}
		\STATE{$\bar{\mathbf{w}}_1 = \mathbf{w}_1 = \mathbf{w}_0$, 
			$\bar{\mathbf{V}}_1 = \mathbf{V}_1 = \mathbf{V}_0$,
			$c_1 = \bar{h}(\mathbf{w}_1, \mathbf{V}_1)$; $q_1 = 1$, $\delta = 10^{-3}$, $a_0 = 0$, $a_1=1$, step-size $\alpha$;}
		\FOR{$t = 1, 2, \dots, T$}
		\STATE
		{$\mathbf{y}_t = \textbf{w}_t + \nicefrac{a_{t-1}}{a_t}(\bar{\mathbf{w}}_t - \textbf{w}_t) + \nicefrac{a_{t-1} - 1}{a_t}(\textbf{w}_t - \textbf{w}_{t-1})$;}
		
		{$\!\mathbf{Y}_t \!= \textbf{V}_t + \nicefrac{a_{t-1}}{a_t}(\bar{\mathbf{V}}_t - \textbf{V}_t) + \nicefrac{a_{t-1} - 1}{a_t}(\textbf{V}_t - \textbf{V}_{t-1})$;}
		
		\STATE
		{$\bar{\textbf{w}}_{t+1} = \Px{\alpha \kappa_0 \hat{\lambda}\hat{\phi}}{\textbf{w}_t - \alpha \nabla_{\mathbf{w}} \bar{\ell}(\textbf{w}_t, \textbf{V}_t)}$;}
		
		{$\bar{\textbf{V}}_{t+1} = \! \Px{\alpha \kappa_0 \bar{\lambda}\bar{\phi}}{\textbf{V}_t - \alpha \nabla_{\mathbf{V}} \bar{\ell}(\textbf{w}_t, \textbf{V}_t)}$;}
		
		\STATE $\Delta_t =  
		\| \bar{\textbf{w}}_{t+1} - \mathbf{y}_t \|_2^2 
		+ \| \bar{\textbf{V}}_{t+1} - \mathbf{Y}_t \|_F^2$
		\IF{$\bar{h}(\bar{\textbf{w}}_{t+1}, \bar{\textbf{V}}_{t+1}) \leq c_t - \delta \Delta_t$;}
		\STATE{$\textbf{w}_{t+1} = \bar{\textbf{w}}_{t+1}$, $\textbf{V}_{t+1} = \bar{\textbf{V}}_{t+1}$;}
		\ELSE
		
		\STATE
		{$\hat{\textbf{w}}_{t+1} = \Px{\alpha \kappa_0 \hat{\lambda}\hat{\phi}}{\textbf{w}_t - \alpha \nabla_{\mathbf{w}}
				\bar{\ell}(\textbf{w}_t, \textbf{V}_t)}$;}
		
		{$\hat{\textbf{V}}_{t+1} = \Px{\alpha \kappa_0 \bar{\lambda}\bar{\phi}}{\textbf{V}_t - \alpha \nabla_{\mathbf{V}}
				\bar{\ell}(\textbf{w}_t, \textbf{V}_t)}$;}
		
		\IF{$\bar{h}(\hat{\textbf{w}}_{t+1}, \hat{\textbf{V}}_{t+1}) < \bar{h}(\bar{\textbf{w}}_{t+1}, \bar{\textbf{V}}_{t+1})$}
		\STATE{$\textbf{w}_{t+1} = \hat{\textbf{w}}_{t+1}$, $\textbf{V}_{t+1} = \hat{\textbf{V}}_{t+1}$;}
		\ELSE
		\STATE{$\textbf{w}_{t+1} = \bar{\textbf{w}}_{t+1}$, $\textbf{V}_{t+1} = \bar{\textbf{V}}_{t+1}$;}
		\ENDIF
		\ENDIF
		\STATE{$a_{t+1} = \nicefrac{1}{2}(\sqrt{4 a_t^2 + 1} + 1)$;}
		\STATE{$q_{t+1} = \eta q_t + 1$, 
			$c_{t+1} = \nicefrac{1}{q_{t+1}} (\eta q_t c_t + \bar{h}(\textbf{w}_{t+1}, \textbf{V}_{t+1}))$;}
		\ENDFOR
		\RETURN{$\textbf{w}_{T+1},\textbf{V}_{T+1}$.}
	\end{algorithmic}
	\label{alg-nmAPG}
\end{algorithm}

\subsubsection{Using SVRG Algorithm}
\label{sec:svrg}

While nmAPG can be an efficient algorithm for \eqref{eq:equiv},
it is still a batch-gradient based method,
which may not be efficient when the sample size is large.
In this case,
the stochastic gradient descent (SGD) \cite{bertsekas1999nonlinear} algorithm is preferred
as it can incrementally update the learning parameters.
However,
the gradient in SGD is very noisy.
To ensure the convergence of SGD,
a decreasing step size must be used,
making the speed possibly even slower than batch-gradient methods.

Recently,
the stochastic variance reduction gradient (SVRG) \cite{xiao2014proximal} algorithm has been developed.
It avoids diminishing step size by introducing variance reduced techniques into gradient updates.
As a result, it combines the best of both worlds,
i.e.,
incremental update of the learning parameters while keeping non-diminishing step size, to achieve significantly faster converging speed than SGD.
Besides, 
it is also extended for the problem in \eqref{eq:compopt} with nonconvex objectives \cite{reddi2016stochastic,allen2016variance}.
This allows the loss function to be smooth (possibly nonconvex) but
the regularizer still needs to be convex.
Thus,
instead of working on the original problem \eqref{eq-obj},
we work on the transformed problem in \eqref{eq:equiv}.

To use SVRG,
we first define the augmented loss for the $n$-th sample as
$\bar{\ell}_{n} (\mathbf{w}, \mathbf{V}) = 
(y^n - \hat{y}^n(\mathbf{w}, \mathbf{V}))^2  
+ \frac{1}{N} g (\mathbf{w}, \mathbf{V})$.
The whole procedure is depicted in Algorithm~\ref{alg:svrg}.
A full gradient is computed in step~4,
a mini-batch $\mathcal{B}$ of size $m_b$ is constructed in step~6, and 
the variance reduced gradient is computed in step~7.
Finally,
the proximal steps can be separately executed based on \eqref{eq:proxsep} in step~8.
As mentioned above, 
the nonconvex variant of SVRG \cite{reddi2016stochastic,allen2016variance} cannot be directly applied to \eqref{eq-obj}.
Instead, we apply it to the transformed problem \eqref{eq:equiv},
where the regularizer becomes convex and the augmented loss is still smooth.
Thus, Algorithm~\ref{alg:svrg} is guaranteed to generate a critical point of \eqref{eq:equiv}.

\begin{algorithm}[ht]
\caption{SVRG for \eqref{eq:equiv}.} 
\small
\begin{algorithmic}[1]
\STATE{Initiate $\bar{\textbf{w}}_0, \bar{\textbf{V}}_0$ as Gaussian random matrices,
	mini-batch size $m_b$;}
\STATE $\mathbf{w}_{1}^B = \bar{\textbf{w}}_0$, $\mathbf{V}_{1}^B = \bar{\textbf{V}}_0$ and step-size $\alpha$;
\FOR{$t = 1, 2, \dots, T$}
	\STATE $\mathbf{w}_{t + 1}^0 = \mathbf{w}_{t}^B$, $\mathbf{V}_{t + 1}^0 = \mathbf{V}_{t}^B$;
	\STATE 
	$\bar{\mathbf{g}}^{ \mathbf{w} }_{t + 1} = \nabla_{ \mathbf{w} } \bar{\ell}(\bar{\mathbf{w}}_t, \bar{\mathbf{V}}_t)$,
	$\bar{\mathbf{g}}^{ \mathbf{V} }_{t + 1} = \nabla_{ \mathbf{V} } \bar{\ell}(\bar{\mathbf{w}}_t, \tilde{\mathbf{V}_t})$;
	
	\FOR{$b = 0, 1, \dots, B - 1$}
		\STATE Uniformly randomly sample a mini-batch $\mathcal{B}$ of size $m_b$;
		
		\STATE 
		$\mathbf{m}_{ \mathbf{w} }^{b}$ $= \! \frac{1}{m_b} 
		\sum_{i_b \in \mathcal{B}}  
		( 
		\nabla_{ \mathbf{w} } \bar{\ell}_{i_b}(\mathbf{w}_t^b, \mathbf{V}_t^b) 
		\! - \!
		\nabla_{ \mathbf{w} } \bar{\ell}_{i_b}(\bar{\mathbf{w}}_t, \bar{\mathbf{V}}_t)
		) 
		\! + \! \bar{\mathbf{g}}^{ \mathbf{w} }_{t + 1}$,
		\\
		$\mathbf{m}_{ \mathbf{V} }^{b} = \frac{1}{m_b} 
		\sum_{i_b \in \mathcal{B}}  
		( 
		\nabla_{ \mathbf{V} } \bar{\ell}_{i_b}(\mathbf{w}_t^b, \mathbf{V}_t^b) 
		- 
		\nabla_{ \mathbf{V} } \bar{\ell}_{i_b}(\bar{\mathbf{w}}_t, \bar{\mathbf{V}}_t)
		) 
		+ \bar{\mathbf{g}}^{ \mathbf{V} }_{t + 1}$;
		
		\STATE 
		$\mathbf{w}_{t + 1}^{b + 1} = \Px{\alpha \kappa_0 \hat{\lambda}\hat{\phi}}{ \mathbf{w}_{t + 1}^b - \alpha \mathbf{m}_{ \mathbf{w} }^{b} }$, 
		
		$\mathbf{V}_{t + 1}^{b + 1} = \Px{\alpha \kappa_0 \bar{\lambda}\bar{\phi}}{ \mathbf{V}_{t + 1}^b - \alpha \mathbf{m}_{ \mathbf{V} }^{b} }$;
	\ENDFOR
	
	\STATE 
	$\bar{\mathbf{w}}_{t + 1} = \frac{1}{B} \sum_{b = 1}^B \mathbf{w}_{t + 1}^b$,
	$\bar{\mathbf{V}}_{t + 1} = \frac{1}{B} \sum_{b = 1}^B \mathbf{V}_{t + 1}^b$;
\ENDFOR
\RETURN{$\bar{\textbf{w}}_{T+1}$, $\bar{\textbf{V}}_{T+1}$.}
\end{algorithmic}
\label{alg:svrg}
\end{algorithm}

\subsubsection{Complexity Analysis}
\label{sec:timecomp}
 
For nmAPG in Algorithm~\ref{alg-nmAPG},
the main computation cost is incurred in performing the proximal steps (step~5 and 10) which cost $O( N K d )$;
then the evaluation of function value (step~7 and 11) costs $O( N K d )$ time.
Thus, the per-iteration time complexity for Algorithm~\ref{alg-nmAPG} is $O( N K d )$.
For SVRG in Algorithm~\ref{alg:svrg},
the computation of the full gradient takes $O( N K d )$ in step~5;
then $O( m_b B K d )$ time is needed for steps~6-10 to perform mini-batch updates.
Thus, one iteration in Algorithm~\ref{alg-nmAPG} takes $O( (N + m_b B) K d )$ time.
Usually,
$m_b B$ shares the same order as $N$ \cite{xiao2014proximal,reddi2016stochastic,allen2016variance}.
Thus, we set $m_b B = N$ in our experiments. As a result,
SVRG needs more time to perform one iteration than nmAPG.
However, due to stochastic updates,
SVRG empirically converges much faster as shown in Section~\ref{sec-exp-svrg-vs-nmAPG}. 

\section{MetaGraph based feature fusion with a hierarchical attention network}
\label{sec-haf}
\begin{figure}[ht]
	\centering
	\includegraphics[width=0.75\textwidth]{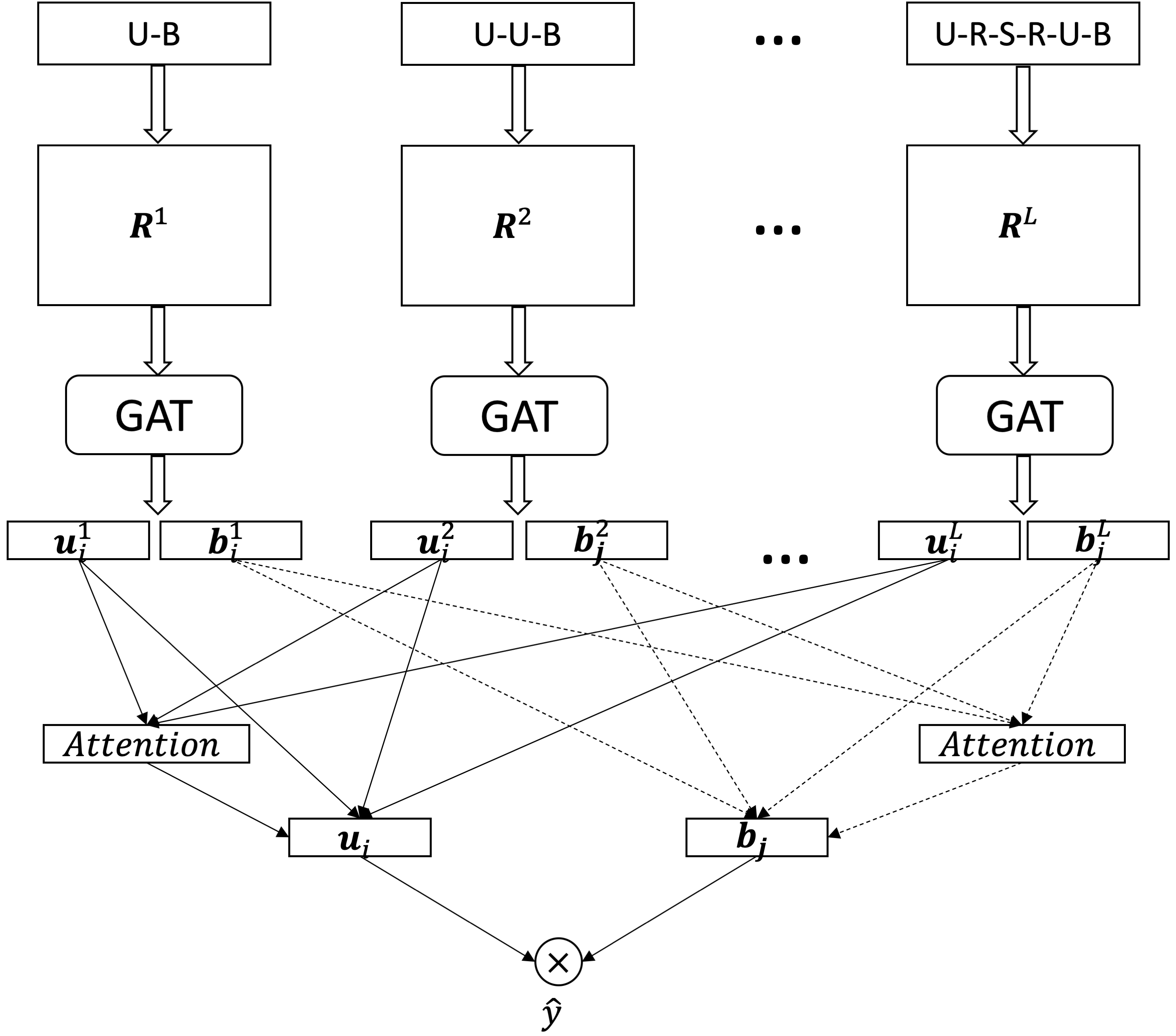}
	\vspace{-10px}
	\caption{HAF is a hierarchical attention model. For each metagraph based user-item similarity matrix, a multi-head GAT is applied to learning embeddings of users and items for each metagraph. To further fuse all metagraph based embeddings, a neural attentive method is introduced. Finally, the predicted rating of user $i$ to business $j$ is computed as the inner product of the final embeddings of $\bu_i$ and $\bb_j$.}
	\label{fig-haf}
\end{figure}

In this section, we propose a novel method based on GNN to fuse metagraph based similarities between users and items.
The general idea is to treat each metagraph based similarity matrix as a new graph. Thus, the representation of each node (user/item) can be learned by adaptively aggregating neighbors' information, 
which is the key component of the recent Graph Neural Networks (GNN) models~\cite{battaglia2018relational}. 
By fusing user and item embeddings from different metagraphs, a neural attention network is employed. 
Therefore, this proposed method is hierarchical, which is dubbed Hierarchical Attention Fusion (HAF).
The architecture is given in Figure~\ref{fig-haf}, 
which can be regarded as a neural version of the ``MF + FM'' framework in Figure~\ref{fig-framework}, 
for its adoption of an end-to-end training manner based on the neural network model.

Note that the architecture of HAF is similar to HAN~\cite{wang2019heterogeneous}, which is also a hierarchical attention method to learn node representation based on metapath in HIN. However, HAF is different from HAN in two aspects. First, HAF tries to learn representation for nodes of different types, i.e., users and items, while HAN is to learn representation for nodes of the same type. Second, metagraph based similarities are used in HAF while in HAN metapath based similarities are used. In the experiments in Section~\ref{sec-exp-rmse}, we create a variant of HAF which only utilizes metapath based similarities. Thus, it can be regarded as the application of HAN to recommendation.

To be specific, by designing $L$ metagraphs, we can get $L$ different user-item similarity matrices, denoted by $\textbf{R}^1, \ldots, \textbf{R}^L$. Following the definition in HAN \cite{wang2019exploring}, we can define metagraph based neighbors in the following:

\begin{definition}[Metagraph based Neighbors]
	Given a node $i$, and a metagraph $M_l$, the metagraph based neighbors $N^l(i)$ of the node $i$ are defined as the set of nodes which connect to node $i$ by instances of the metagraph $M_l$.
\end{definition}

Then the $L$ metagraph based similarity matrices can be regarded as adjacency matrices for $L$ bipartite graphs. Thus, the metagraph based neighbors of a user are items, while the metagraph based neighbors of an item are users. 
We then elaborate on the technical details about HAF for the rating prediction task.

\subsection{Metagraph-based User and Item Embedding Learning}
Since user and item are of different node types, we design type-specific encoders for users and items, respectively, which transform them into the same embedding space. To be specific, for a user $i$, 
\begin{equation}
\bh_i^l = \bW_{\phi_i} \cdot \be_i^l,
\label{eq-type-encoding}
\end{equation}
where $\be_i^l$ and $\bh_i^l$, respectively, are the original and transformed features of user $i$ from the $l$-th bipartite graph, and $\bW_{\phi_i}$ is the transformation matrix for user node type $\phi_i$. For simplicity, we remove the superscript $l$ when there is no misunderstanding. Since the processing of users and items are the same, we use $\bh_j$ and $\be_j$ to represent the original and transformed features given an item $j$.

Based on Eq.~\eqref{eq-type-encoding}, we can obtain $L$ groups of transformed user and item features. Next, we show how to learn user and item representations by aggregating these transformed features following graph attention network (GAT) \cite{velivckovic2017graph}, i.e., using the self-attention mechanism when learning user and item embeddings.


For a user-item pair $(i, j)$ from the $l$-th metagraph based similarity matrix, the node-level attention, i.e., self-attention, $\bs_{ij}$ represents the importance of item $j$ to user $i$, 
which can be computed as $s_{ij} = Att(\bh_i, \bh_j)$, 
where $Att$ is typically a neural network function to compute the attention coefficients. 
Then $s_{ij}$ is normalized by softmax function as follows:
\begin{align}
a_{ij} = \text{softmax}_j(s_{ij}) 
= 
\exp\big(\sigma(\ba^{\top} \cdot [\bh_i||\bh_j])\big) / \sum\nolimits_{k \in N^l(i)} \exp\big(\sigma(\ba^{\top} \cdot [\bh_i||\bh_k])\big),
\label{eq-self-attention}
\end{align}
where $\sigma$ is the activation function, e.g., LeakyReLU, and $||$ represents the concatenation operation. $\ba \in \mathbb{R}^{2F}$ is the parameterized weight vector. Then the final representation of node $i$ is computed as $\bz_i = \sigma\big(\sum_{j \in N(i)}a_{ij} \cdot \bh_j\big)$. Note that for simplicity, we remove the superscript of each notation, representing the corresponding metagraph, except for metagraph based neighbors $N^l(i)$.

To enhance the robustness, we use  the multi-head attention, which repeats the node-level attention $K$ times, and concatenates the learnt embeddings as the final representation:
\begin{equation}
\bz_i = \overset{K}{\underset{k=1}{||}}\sigma\big(\sum\nolimits_{j \in N(i)}a_{ij} \cdot \bh_j\big).
\label{eq-multi-head}
\end{equation}
Eq.~\eqref{eq-self-attention} and \eqref{eq-multi-head} give the computation process for obtaining the embeddings of user $i$ from a given megagraph $M_l$, and for item $j$, we can obtain its embeddings following the same process. Thus, for the user-item pair $(i, j$), we can obtain $L$ groups of embeddings by the node-level attention, denoted as $\bu^1_i, \cdots, \bu^L_i$, $\bb^1_j, \cdots, \bb^L_j$.

\subsection{Metagraph based User and Item Embedding Fusion}
As introduced aforementioned, different metagraphs capture similarities of different semantics, and the $L$ groups of embeddings represent different aspects of each user and item. 
Thus, we need to fuse all metagraph based embeddings of users and items. Intuitively, the importance of each embedding should be different for the final representation. 
To capture this, we further design another attention mechanism to aggregate embeddings from different metagraphs adaptively. 
For user $i$, given $L$ groups of embeddings, $\bu^1_i, \cdots, \bu^L_i$, the attention is computed as 
\begin{equation}
p_l = \frac{1}{|\mathcal{V}|} 
\sum\nolimits_{i \in \mathcal{V}} \bq^{\top} 
\!\!\!
\cdot \text{tanh}(\bW_u \cdot \bu_i^l + \mathbf{b_u}),
\label{eq-metagraph-attention}
\end{equation}
where $\bW_u$ is the weight matrix, $\bb_u$ is the bias vector, and $\bq$ is the metagraph-level attention vector. All of the above parameters are shared for all metagraphs for meaningful comparisons.
Likewise, we normalize $p_l$ over all metagraphs by softmax function $g_l = \nicefrac{\exp(p_l)}{\sum_{l = 1}^{L} \exp(p_l)}$. 
With these learned attention weights, 
we can fuse all metagraph based embeddings to obtain the final representation for user $i$ as follows:
\begin{equation*}
\bu_i = \sum\nolimits_{l= 1}^{L} g_l \cdot \bu_i^l.
\end{equation*}
The final representation for item $j$ can be obtained in the same process, denoted as $\bb_j$.
After obtaining the final representation for user $i$ and item $j$, 
the predicted rating is computed by dot production, $\hat{y} = \bu_i^{\top} \bb_j$. 
To train the whole framework, we use the mean square loss in Eq.~\eqref{eq:loss} and obtain task-specific user and item representations via back propagation under the supervision of the observed ratings.

\subsection{Complexity Analysis}
HAF is highly efficient since the computation of node-level and metagraph-level attention can be implemented in parallel for each node and each metagraph. Similar to GAT, the complexity of HAF is $O( (|\mathcal{V}_u| + |\mathcal{V}_b|) \cdot F_e \cdot F_h \cdot K + |\mathcal{E}_{ub}| \cdot F_h \cdot K + (|\mathcal{V}_u| + |\mathcal{V}_b|) \cdot F_q \cdot F_h \cdot K )$, where $|\mathcal{V}_u|$ and $\mathcal{V}_b$ are the number of users and businesses, respectively. $K$ is the number of heads in Eq.~\eqref{eq-multi-head}. $\mathcal{E}_{ub}$ is the number of non-zero elements in each metagraph based similarity matrix.
We remove the superscript denoting the specific metagraph for simplicity, 
since the computation process can be executed individually. $F_e, F_h, F_q$ are the dimension size of the vector $\be, \bh, \bq$, respectively.

\section{Experiments}
\label{sec-exp}

In this section, we conduct extensive experiments to demonstrate the effectiveness of our proposed frameworks, including FMG and HAF. We first introduce the datasets, evaluation metrics and experimental settings in Section~\ref{sec-exp-setting}. In Section~\ref{sec-exp-rmse}, we show the recommendation performance of our proposed frameworks compared to several state-of-the-art recommendation methods.
We analyze the influence of the both convex and nonconvex group lasso regularization in Section~\ref{sec-exp-vary-reg}. 
Besides the group lasso regularizers, in Section~\ref{sec-aspect-fmg}, we further analyze other important components of FMG, including the feature extraction methods in the MF part, the influence of the rank of second-order weight matrix in the FM part, and comparisons between different optimization algorithms in the FM part.  
Further, we give a detailed analysis of different parameters of HAF framework as well as visualize the attention weight distribution over metagraphs in Section~\ref{sec-exp-analysis-haf}.

\subsection{Setup}
\label{sec-exp-setting}

To demonstrate the effectiveness of HIN for recommendation, we conduct experiments using two datasets with rich side information. The first dataset is Yelp, which is provided for the Yelp challenge.\footnote{https://www.yelp.com/dataset\_challenge} Yelp is a website where a user can rate local businesses or post photos and review about them. The ratings fall in the range of 1 to 5, where higher ratings mean users like the businesses while lower rates mean users dislike business. Based on the collected information, the website can recommend businesses according to the users' preferences. The second dataset is Amazon Electronics,\footnote{http://jmcauley.ucsd.edu/data/amazon/} which is provided in~\cite{he2016ups}. As we know, Amazon highly relies on RSs to present interesting items to its users. 
We extract subsets of entities from Yelp and Amazon to build the HINs, which include diverse types and relations. The subsets of the two datasets both include around 200,000 ratings in the user-item rating matrices. Thus, we identify them as Yelp-200K and Amazon-200K, respectively. In addition, to better compare our frameworks with existing HIN-based methods, we also use the datasets provided in the CIKM paper~\cite{shi2015semantic}, which are denoted as CIKM-Yelp and CIKM-Douban, respectively. Note that four datasets are used to compare the recommendation performance of different methods, as shown in Section~\ref{sec-exp-rmse}. To evaluate other aspects of our model, we only conduct experiments on the first two datasets, i.e., the Yelp-200K and Amazon-200K datasets. The statistics of our datasets are shown in Table~\ref{tb-stat}. For the detailed information of CIKM-Yelp and CIKM-Douban, we refer the readers to~\cite{shi2015semantic}. 

\begin{table}[ht]
\centering
\caption{Statistics of the Yelp-200K and Amazon-200K datasets.} 
\vspace{-10px}
\small
\label{tb-stat}
\begin{tabular}{c|c|cccc c}
	\hline
	& Relations(A-B) & \begin{tabular}[c]{@{}c@{}}
		Number \\
		of A
	\end{tabular} & \begin{tabular}[c]{@{}c@{}}
		Number \\
		of B
	\end{tabular} & \begin{tabular}[c]{@{}c@{}}
	 Number  \\
		of (A-B)
	\end{tabular} & \begin{tabular}[c]{@{}c@{}}
		Avg Degrees \\
		  of A/B
	\end{tabular} \\ \hline
 \multirow{5}{*}{Amazon-200K}	&User-Review & 59,297 & 183,807 & 183,807 & 3.1/1 \\ \cline{2-6}
 &	Business-Category & 20,216 & 682 & 87,587 & 4.3/128.4 \\ \cline{2-6}
 &	Business-Brand & 95,33 & 2,015 & 9,533 & 1/4.7 \\ \cline{2-6}
 &	Review-Business & 183,807 & 20,216 & 183,807 & 1/9.1 \\ \cline{2-6}
 &	Review-Aspect & 183,807 & 10 & 796,392 & 4.3/79,639.2 \\
 \hline\hline
\multirow{9}{*}{Yelp-200K}	& User-Business & 36,105 & 22,496 & 191,506 & 5.3/8.5 \\ \cline{2-6}
	&User-Review & 36,105 & 191,506 & 191,506 & 5.3/1 \\ \cline{2-6}
&User-User & 17,065 & 17,065 & 140,344 & 8.2/8.2 \\ \cline{2-6}
&Business-Category & 22,496 & 869 & 67,940 & 3/78.2 \\ \cline{2-6}
&Business-Star & 22,496 & 9 & 22,496 & 1/2,499.6 \\ \cline{2-6}
&Business-State & 22,496 & 18 & 22496 & 1/1,249.8 \\ \cline{2-6}
&Business-City & 22,496 & 215 & 22,496 & 1/104.6 \\ \cline{2-6}
&Review-Business & 191,506 & 22,496 & 191,506 & 1/8.5 \\ \cline{2-6}
&Review-Aspect & 191,506 & 10 & 955,041 & 5/95,504.1 \\
\hline
\end{tabular}	
\end{table}

\begin{table}[ht]
	\centering
	\small
	\caption{The density of rating matrices in the four datasets (
		{\normalfont $\text{Density}=\frac{\text{\#Ratings}}{\text{\#Users}\times \text{\#Items}}$}).} 
	\vspace{-10px}
	\label{tb-exp-data-density}
	\begin{tabular}{c| cccc}
		\hline
		               & Amazon-200K & Yelp-200K & CIKM-Yelp & CIKM-Douban \\ \hline
		\text{Density} & 0.015\%     & 0.024\%   & 0.086\%   & 0.630\%     \\ \hline
	\end{tabular}
\end{table}	

To evaluate the recommendation performance, we adopt the root-mean-square-error (RMSE) as our metric, which is the most popular for rating prediction in the
literature~\cite{koren2008factorization,ma2011recommender,mnih2007probabilistic}. 
It is defined as
$\text{RMSE} = 
\sqrt{\sum_{y^n \in \mathcal{R}_{test}} (y^n - \hat{y}^n)^2}
/
\sqrt{|\mathcal{R}_{test}|}$,
where $\mathcal{R}_{test}$ is the set of all the test samples, $\hat{y}^n$ is the predicted rating for the $n$-th sample, $y^n$ is the observed rating of the $n$-th sample in the test set. A smaller RMSE value means better performance.

We compare the following baseline models to our approaches.
\begin{itemize}[leftmargin=5mm]
\item \textbf{RegSVD}~\cite{paterek2007improving}: The basic matrix factorization model with $L_2$ regularization, 
which uses only the user-item rating matrix. We use the implementation in~\cite{guo2015librec}.

\item \textbf{FMR}~\cite{rendle2012fm}: The factorization machine with only the user-item rating matrix. We adopt the method in Section 4.1.1 of~\cite{rendle2012fm} to model the rating prediction task. We use the code provided by the authors.\footnote{\url{http://www.libfm.org/}}

\item \textbf{HeteRec}~\cite{yu2014personalized}: It is based on metapath based similarity between users and items. A weighted ensemble model is learned from the latent features of users and items generated by applying matrix factorization to the similarity matrices of different metapaths. We implemented it based on \cite{yu2014personalized}.

\item \textbf{SemRec}~\cite{shi2015semantic}: It is a metapath based recommendation technique on weighted HIN, which is built by connecting users and items with the same ratings. Different models are learned from different metapaths, and a weight ensemble method is used to predict the users' ratings. We use the code provided by the authors.\footnote{\url{https://github.com/zzqsmall/SemRec}}

\item \textbf{FMG}: The proposed framework (Figure~\ref{fig-framework}) with convex group lasso regularizer in \eqref{eq-glasso-w} and \eqref{eq-glasso-v} used with factorization machine. The model is proposed in our previous work~\cite{zhao2017meta}; 
\textbf{FMG(LSP)}: Same as FMG, except nonconvex group lasso regularizer in \eqref{eq:gpreg} is used.

\item \textbf{HAF(MP)}: the proposed hierarchical graph attention network with metapath, i.e., on Yelp, $M_9$ is removed, and on Amazon, $M_6$ is removed. 

\item \textbf{HAF(MG)}: the proposed hierarchical graph attention network with all metagraphs.
\end{itemize}
Note that it is reported in~\cite{shi2015semantic} that SemRec outperforms the method in \cite{yu2013collaborative},
which uses metapath based similarities as regularization terms in matrix factorization. 
Thus, we do not include \cite{yu2013collaborative} for comparison. All methods, except for HAF, are run in a server (OS: CentOS release 6.9, CPU: Intel i7-3.4GHz, RAM: 32GB). For HAF, it is run on a GPU 2080Ti (Memory: 12GB, Cuda version: 10.2).
For baselines and FMG methods, they are implement with Python 2.7.5, and for HAF methods, they are implement with Pytorch (version 1.2)~\cite{paszke2019pytorch} on top of Deep Graph Library (DGL) \cite{wang2019dgl}.

On Amazon-200K and Yelp-200K datasets, we use the metagraphs in Figures~\ref{fig-amazon-metagraph} and \ref{fig-yelp-metagraph} for HeteRec, SemRec, FMG, and FMG(LSP), while on CIKM-Yelp and CIKM-Douban, we use the metapaths provided in ~\cite{shi2015semantic} for these four methods. To get the aspects (e.g., $A$ in Figures~\ref{fig-yelp-metagraph} and \ref{fig-amazon-metagraph}) from review texts, 
we use a topic model software Gensim~\cite{rehurek_lrec} to extract topics from the review texts and use the extracted topics as aspects. 
The number of topics is set to $10$ empirically. 

In Section~\ref{sec-exp-rmse}, we use the four datasets in Table~\ref{tb-exp-data-density} to compare the recommendation performance of our models and the baselines. For the experimental settings, we randomly split the whole dataset into 80\% for training, 10\% for validation and the remaining 10\% for testing. The process is repeated five times and the average RMSE of the five rounds is reported. Besides, for the parameters of our models, we set $\hat{\lambda}=\bar{\lambda} = \lambda$ in Eq.~\eqref{eq-obj} for simplicity, and $\lambda$ is set to obtain the optimal value on different validation datasets. As in~\cite{zhao2017meta}, $F$ and $K$ are set to $10$ for its good performance and computational efficiency. 
For the two variants of HAF, the hidden vector size is set to $32$, the output embedding size to $32$, the number of heads in self-attention part to $4$, the learning rate and $L_2$ norm to $0.0001$, the dropout is set to $0.1$, and the training epochs to $200$.
From Sections~\ref{sec-exp-vary-reg} to Section~\ref{sec-exp-analysis-haf}, to explore the influences of different settings of the proposed frameworks, we create two smaller datasets, Amazon-50K and Yelp-50K, where only 50,000 ratings are randomly sampled from Amazon-200K and Yelp-200K. 

\subsection{Recommendation Effectiveness}
\label{sec-exp-rmse}

The RMSEs of all methods are shown in Table~\ref{tb-exp-rmse}.
For CIKM-Yelp and CIKM-Douban, we directly report the performance of SemRec from~\cite{shi2015semantic} since the same amount of training data is used in our experiement.
Besides, the results of SemRec on Amazon-200K are not reported, as the programs crashed due to large demand of memory.

\begin{table}[ht]
\centering
\caption{Recommendation performance of all approaches in terms of RMSE.
	The lowest RMSEs (according to the pairwise t-test with 95\% confidence) are highlighted.}
\vspace{-5px}
\label{tb-exp-rmse}
\small
\begin{tabular}{c|c|c|c|c}
	\hline
	          &        Amazon-200K         &         Yelp-200K          &         CIKM-Yelp          &        CIKM-Douban         \\ \hline
	{RegSVD}  &     2.9656$\pm$0.0008      &     2.5141$\pm$0.0006      &     1.5323$\pm$0.0011      &     0.7673$\pm$0.0010      \\ \hline
	  {FMR}   &     1.3462$\pm$0.0007      &     1.7637$\pm$0.0004      &     1.4342$\pm$0.0009      &     0.7524$\pm$0.0011      \\ \hline
	{HeteRec} &     2.5368$\pm$0.0009      &     2.3475$\pm$0.0005      &     1.4891$\pm$0.0005      &     0.7671$\pm$0.0008      \\ \hline
	{SemRec}  &            ---             &     1.4603$\pm$0.0003      &         1.1559(*)          &         0.7216(*)          \\ \hline\hline
	   FMG    &     1.1953$\pm$0.0008      &     1.2583$\pm$0.0003      &     1.1167$\pm$0.0011      &     0.7023$\pm$0.0011      \\ \hline
	FMG(LSP)  &     1.1980$\pm$0.0010      &     1.2593$\pm$0.0005      &     1.1255$\pm$0.0012      & \textbf{0.7035$\pm$0.0013} \\ \hline
	 HAF(MP)  &     1.1953$\pm$0.0003      &     1.2470$\pm$0.0008      &     1.1170$\pm$0.0012      &     0.7035$\pm$0.0017      \\ \hline
	 HAF(MG)  & \textbf{1.1905$\pm$0.0002} & \textbf{1.2335$\pm$0.0007} & \textbf{1.1139$\pm$0.0010} & \textbf{0.6975$\pm$0.0021} \\ \hline
\end{tabular}
\end{table}


Firstly, we can see that both FMG and FMG(LSP) consistently outperform all baselines on the four datasets. 
This demonstrates the effectiveness of the proposed ``MF + FM'' framework shown in Figure~\ref{fig-framework}. 
Note that the performance of FMG and FMG(LSP) are very close, but
FMG(LSP) needs fewer features to achieve such performance, 
which supports our motivation to use nonconvex regularization for selecting features.
In the following two sections, we will compare in detail the two regularizers. Besides, HAF can further decrease RMSE, which demonstrates the effectiveness of the hierarchical attention mechanism. In other words, it demonstrates the superiority of deep learning methods for HIN-based RS, which have been explored in recent works~\cite{shi2019deep,Hu2018LMB,fan2019metapath,jin2020efficient}.  Note that HAF(MG) outperforms HAF(MP), thus demonstrating the importance of complex semantics captured by metagraphs, which motivates the introduction of metagraph in the conference version of this work \cite{zhao2017meta}.

Secondly, from Table \ref{tb-exp-rmse},
we can see that comparing to RegSVD and FMR, 
which only use the rating matrix, SemRec and FMG, 
which use side information from metagraphs, are significantly better. 
In particular, the sparser the rating matrix, the more significant is the benefit produced by the additional information.
For example, on Amazon-200K, FMG outperforms RegSVD by 60\%, while for CIKM-Douban, the percentage of RMSE decrease is 8.5\%. 
Note that the performance of HeteRec is worse than FMR, despite the fact that we have tried our best to tune the model. 
This aligns with our discussion in Section~\ref{sec-framework-model} that a weighting ensemble of dot products of latent features may cause information loss among the metagraphs and fail to reduce noise caused by having too many metagraphs. These demonstrate the effectiveness of FMG for fusing various side information for recommendation.

When comparing the results of FMG and SemRec, 
we find that the performance gap between them are not that large, 
which means that SemRec is still a good method for rating prediction, 
especially when comparing to the other three baselines. The good performance of SemRec may be attributed to the reason that it incorporates rating values into HIN to create a weighted HIN, which can better capture the metagraph or metapath based similarities between users and items.

\begin{figure}[ht]
	\centering
	\subfigure[Amazon-50K.]{
		\label{fig-amazon-rmse-vary-lambda}
		\includegraphics[width=0.40\textwidth]{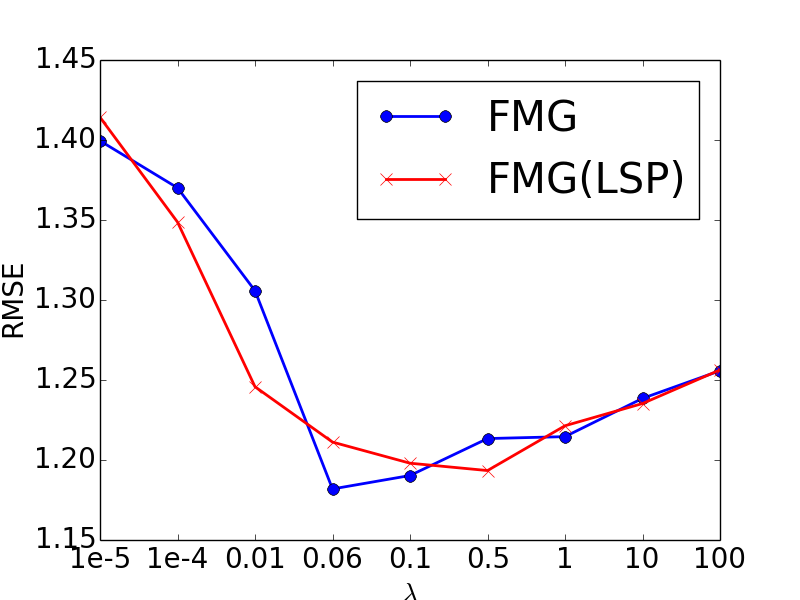}}	
	\subfigure[Yelp-50K.]{
		\label{fig-yelp-rmse-vary-lambda}
		\includegraphics[width=0.40\textwidth]{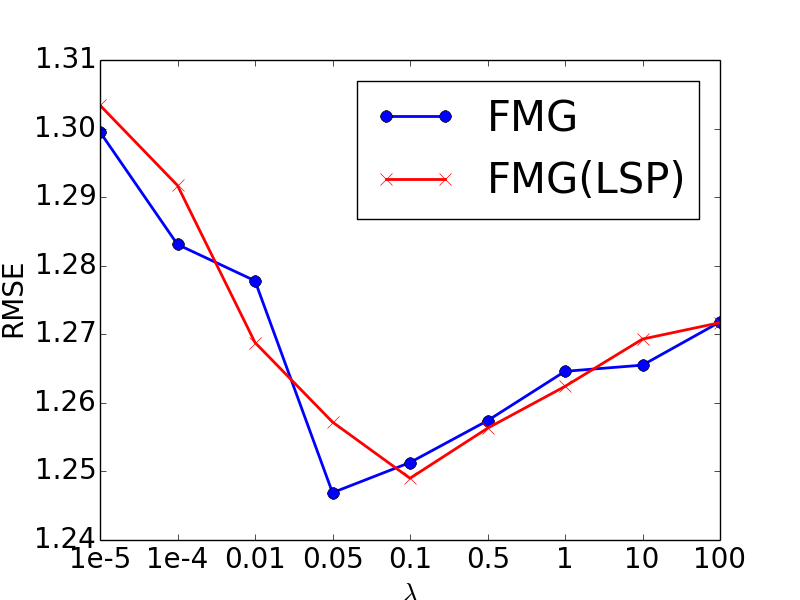}}
	\vspace{-10px}
	\caption{RMSE v.s $\lambda$ on the Amazon-50K and Yelp-50K datasets.}
	\centering
	\label{fig-rmse-vary-lambda}
\end{figure}

\subsection{The Impact of Group Lasso Regularizer}
\label{sec-exp-vary-reg}

In this part, we study the impact of group lasso regularizer for FMG.
Specifically, we show the trend of RMSE by varying $\lambda$
(with $\hat{\lambda} = \bar{\lambda} = \lambda$ in~\eqref{eq-obj}), 
which controls the weights of group lasso.
 The RMSEs of Amazon-50K and Yelp-50K are shown in
Figure~\ref{fig-rmse-vary-lambda}(a) and (b), respectively. 
We can see that with $\lambda$ increasing, RMSE decreases first and then increases, 
demonstrating that $\lambda$ values that are too large or too small are not good for the performance of rating prediction. 
Besides,
we observe that the trend of the FMG(LSP) is similar to that of FMG,
except that the best performance is achieved with different $\lambda$'s.

\begin{figure}[ht]
	\centering
	\subfigure[Amazon-50K.]
	{\includegraphics[width=0.40\textwidth]{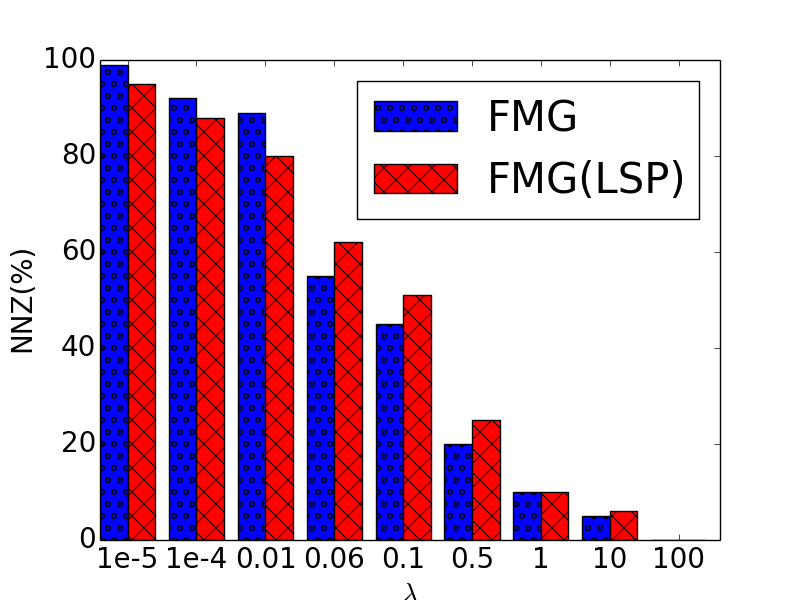}}
	\subfigure[Yelp-50K.]
	{\includegraphics[width=0.40\textwidth]{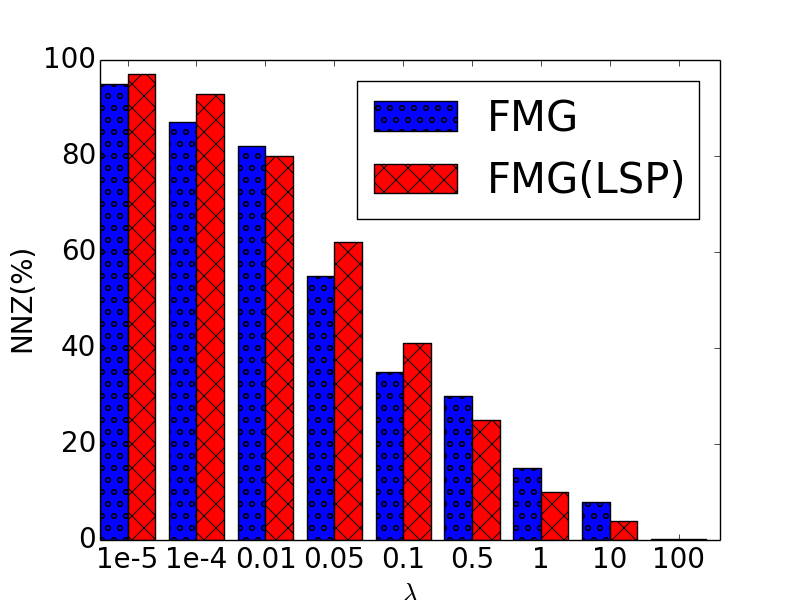}}
	\vspace{-10px}
	\caption{The trend of NNZ by varying $\lambda$ on the Amazon-50K and Yelp-50K datasets. On Amazon-50K, FMG performs best when $\lambda = 0.06$, and FMG(LSP) is the best when $\lambda = 0.5$. On Yelp-50K, FMG performs best when $\lambda = 0.05$, and FMG(LSP) best when $\lambda = 0.1$.}
	\label{fig-sparsity}
\end{figure}

\subsubsection{Sparsity of $\textbf{w},\textbf{V}$}
\label{sec-sparsity-fmg}
We study the sparsity of the learned parameters, i.e., the ratio of zeros in $\textbf{w},\textbf{V}$, after learning.
We define NNZ (number of non zeros) as $\frac{nnz}{w_n+v_n}$,
where $nnz$ is the total number of nonzero elements in $\textbf{w}$ and $\textbf{V}$, and $w_n$ and $v_n$ are the number of entries in $\textbf{w}$ and $\textbf{V}$, respectively. The smaller NNZ, the fewer the nonzero elements in $\textbf{w}$ and $\textbf{V}$, and the fewer the metagraph based features left after training. The trend of NNZ with different $\lambda$'s is shown in
Figure~\ref{fig-sparsity}. We can see that with $\lambda$ increasing, 
NNZ becomes smaller, which aligns with the effect of group lasso. 
Note that the trend is non-monotonic due to the nonconvexity of the objective.
Besides, 
NNZ of the parameters of FMG(LSP) is much smaller than that of FMG when the best performance on both Amazon-50K and Yelp-50K is achieved. This is due to the effect of nonconvexity of LSP, which can induce larger sparsity of the parameters with a smaller loss of performance gain.

Besides, we emphasize an interesting discovery here. 
From Figure~\ref{fig-sparsity}, 
we can see that on both Amazon-50K and Yelp-50K the NNZ of FMG(LSP) is smaller than that of FMG when they both obtain the best performance. 
For example, on Amazon-50K, FMG performs best with $\lambda = 0.06$ and $NNZ=0.52$, 
while FMG(LSP) performs best with $\lambda = 0.5$ and $NNZ=0.25$. 
Similar cases exist on Yelp-50K. 
In other words, to obtain the best performance, 
nonconvex regularizers can induce larger sparsity, 
which means they can select useful features more effectively, 
i.e., they can achieve comparable performance with fewer selected metagraphs.

\subsubsection{The Selected Metagraphs}
\label{sec-sel-graphs-convex}

Recall that in Eq.~\eqref{eq-rating-fm}, we introduce $\textbf{w}$ and $\textbf{V}$, respectively, to capture the first-order weights for the features and second-order weights for interactions of the features. Thus, after training, the nonzero values in $\textbf{w}$ and $\textbf{V}$ represent the selected features, i.e., the selected metagraphs. 
We list in Table~\ref{tb-selected-mg} the selected metagraphs corresponding to nonzero values in $\textbf{w}$ and $\textbf{V}$ from the perspective of both users and items,
when lowest RMSEs are achieved.

\begin{table}[ht]
\centering
\caption{The selected metagraphs by FMG and FMG(LSP) on Amazon-50K and Yelp-50K datasets. 
	We show the selected latent features from the perspective of users and items and from both first-order and seconder-order parameters.}
\vspace{-10px}
\small
\begin{tabular}{c | c | c | c | c | c}
	\hline
	       &            &          \multicolumn{2}{c|}{User-Part}           &        \multicolumn{2}{c}{Item-Part}        \\ \cline{2-6}
	       &            & first-order             & second-order            & first-order             & second-order      \\ \hline
	Amazon & FMG        & $M_1$-$M_3$,$M_5$       & $M_1$-$M_6$             & $M_2$,$M_3$,$M_5$,$M_6$ & $M_2$,$M_5$,$M_6$ \\ \cline{2-6}
	 -50K  & FMG(LSP)   & $M_1$,$M_5$             & -                       & $M_2$,$M_5$             & -                 \\ \hline
	 Yelp  & FMG        & $M_1$-$M_4$,$M_6$,$M_8$ & $M_1$-$M_3$,$M_5$,$M_8$ & $M_1$-$M_5$,$M_8$,$M_9$ & $M_3$,$M_8$       \\ \cline{2-6}
	 -50K  & FMG(LSP)   & $M_1$,$M_3$,$M_4$,$M_8$ & $M_2,M_3,M_8$           & $M_1$-$M_5$,$M_8$       & $M_8$             \\ \hline
\end{tabular}
\label{tb-selected-mg}
\end{table}

From Table~\ref{tb-selected-mg}, 
for FMG,
we can observe that the metagraphs with style like $U\rightarrow *\leftarrow U\rightarrow B$ are better than those like $U\rightarrow B \rightarrow * \leftarrow B$.
We use $U\rightarrow *\leftarrow U\rightarrow B$ to represent metagraphs like $M_2, M_3, M_8, M_9$ in Figure~\ref{fig-yelp-metagraph} (Yelp) and $M_2, M_5, M_6$ in Figure~\ref{fig-amazon-metagraph} (Amazon), and $U\rightarrow B \rightarrow * \leftarrow B$ to represent metagraphs like $M_4, M_5, M_6, M_7$ in Figure~\ref{fig-yelp-metagraph} and $M_3, M_4$ in Figure~\ref{fig-amazon-metagraph}. On Yelp-50K, we can see that metagraphs like $M_2, M_3, M_8, M_9$ tend to be selected while $M_4 - M_7$ are removed. This means that on Yelp, recommendations by friends or similar users are better than those by similar items. Similar observations can be made on Amazon-50K, i.e., $M_3,M_4$ tend to be removed. 
Furthermore, on both datasets, complex structures like $M_9$ in Figure~\ref{fig-yelp-metagraph} and $M_6$ in Figure~\ref{fig-amazon-metagraph} are found to be important for item latent features. This demonstrates the importance of the semantics captered by metagraphs, which are ignored in previous metapath based RSs~\cite{yu2013collaborative,yu2014personalized,shi2015semantic}.
The observations on FMG(LSP) are very similar to that of FMG, 
i.e., metagraphs with style like $U\rightarrow *\leftarrow U\rightarrow B$ are better than those like $U\rightarrow B \rightarrow * \leftarrow B$. On Yelp-50K, metagraphs like $M_2, M_3, M_8$ tend to be selected while $M_4 - M_7$ are removed, while on Amazon-50K $M_3,M_4$ tend to be removed.

\begin{figure}[ht]
	\centering
	\subfigure[Amazon-50K.]{%
		\label{fig-amazon-single-mg}
		\centering
		\includegraphics[width=0.40\textwidth]{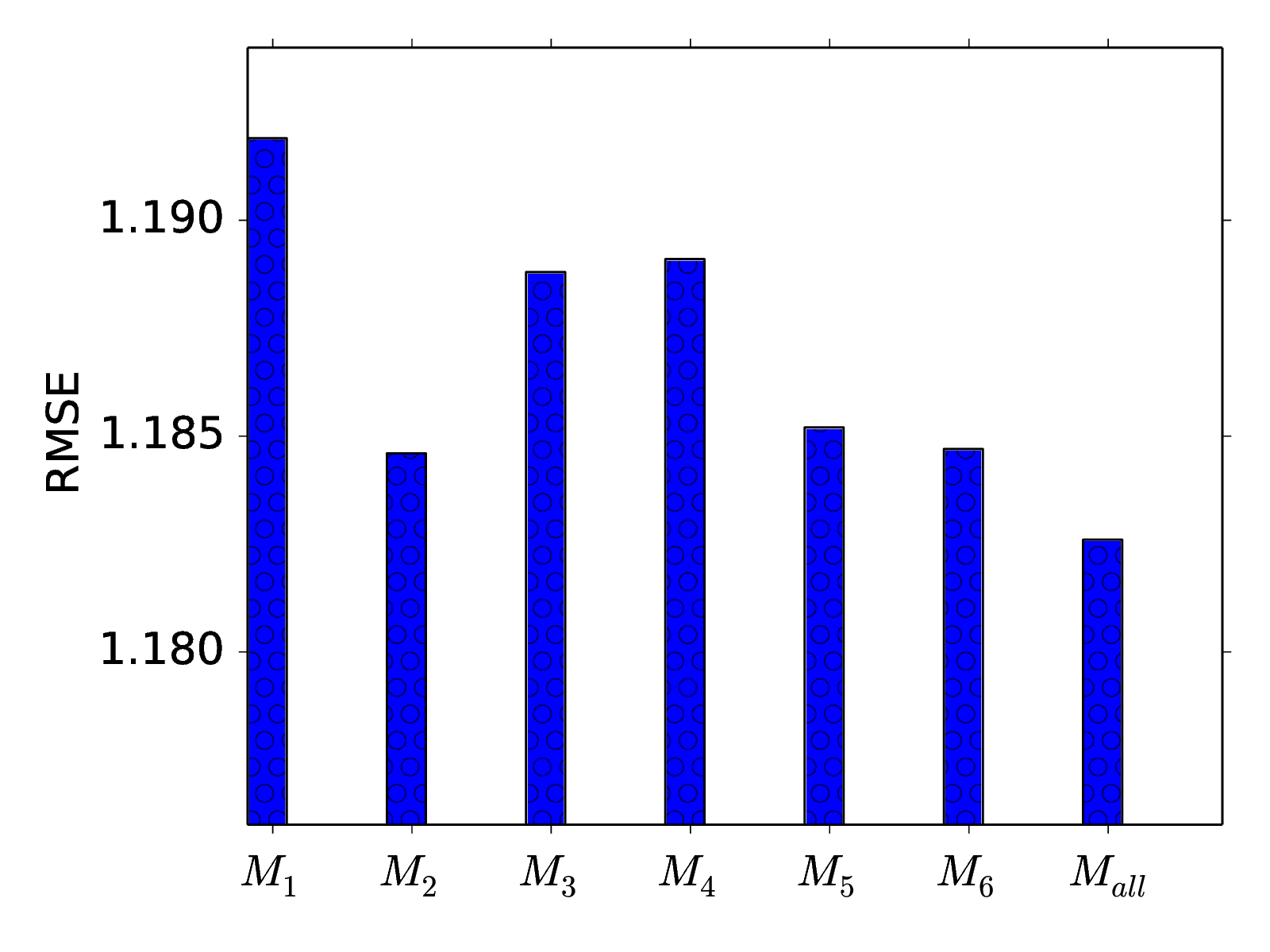}%
	}
	\subfigure[Yelp-50K.]{
		\label{fig-yelp-single-mg}
		\centering
		\includegraphics[width=0.40\textwidth]{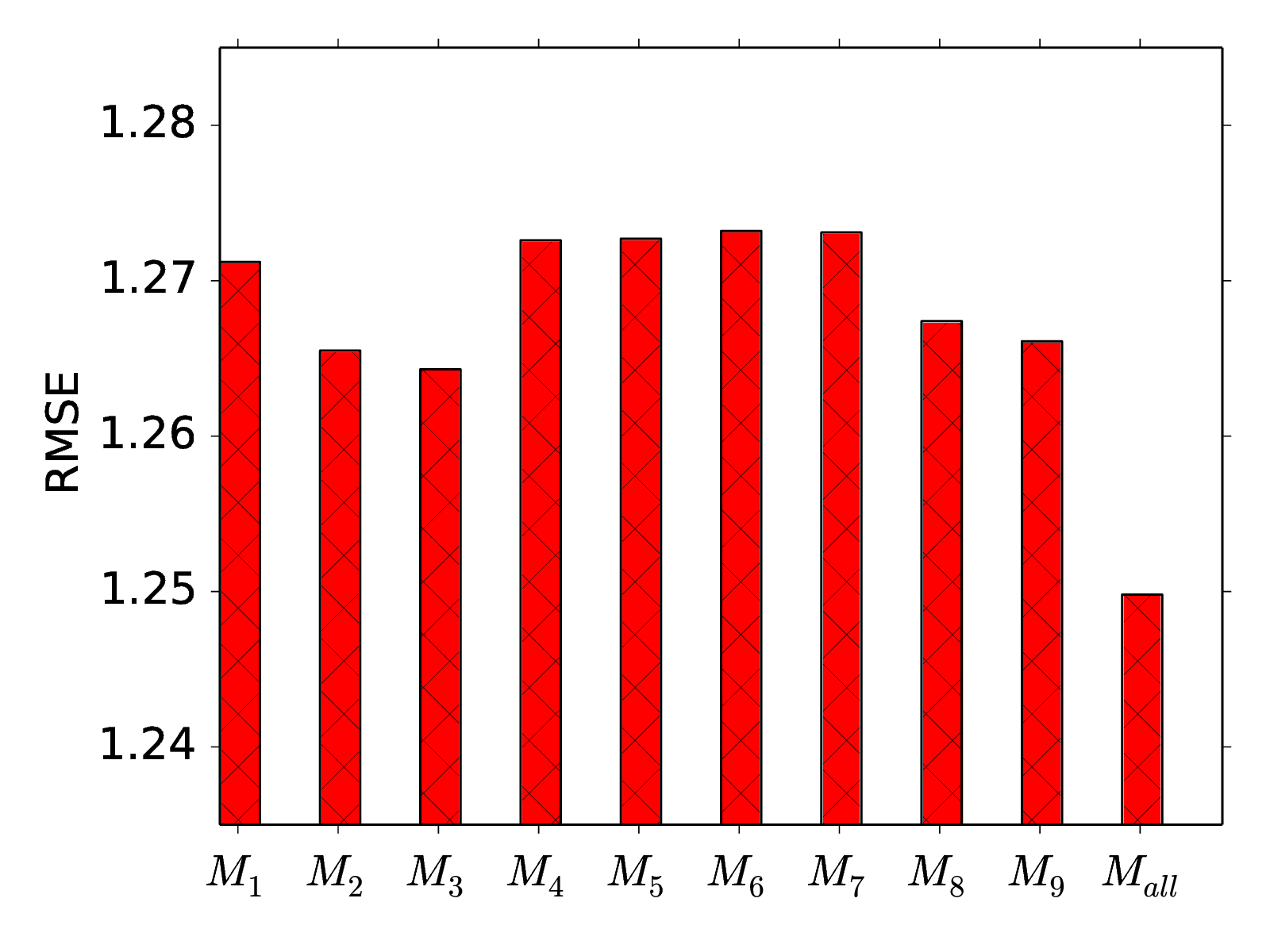}%
	}
	\vspace{-10px}
	\caption{RMSE of each metagraph on the Amazon-50K and Yelp-50K datasets. $M_{all}$ is our model trained with all metagraphs.}
	\label{fig-rmse-single-mg}
\end{figure}

\subsubsection{Performance with Single Metagraph}
\label{sec-exp-single-rmse}

In this part, we compare the performance of different metagraphs separately on Amazon-50K and Yelp-50K. In the training process, we use only one metagraph for user and item features and then predict with FMG and evaluate the results obtained by the corresponding metagraph. Specifically, we run experiments to compare RMSE of each metagraph in Figures~\ref{fig-yelp-metagraph} and \ref{fig-amazon-metagraph}. The RMSE of each metagraph is shown in Figure~\ref{fig-rmse-single-mg}. Note that we show for comparison the RMSE when all metagraphs are used, which is denoted by $M_{all}$.

From Figure~\ref{fig-rmse-single-mg}, we can see that on both Amazon-50K and Yelp-50K, 
the performance is the best when all metagraph-based user and item features are used, 
which demonstrates the usefulness of the semantics captured by the designed metagraphs in Figures~\ref{fig-yelp-metagraph} and \ref{fig-amazon-metagraph}. Besides, we can see that on Yelp-50K, the performance of $M_4-M_7$ is the worst, and on Amazon-50K, the performance of $M_3-M_4$ is also among the worst three. Note that they are both metagraphs with style like $U\rightarrow B \rightarrow * \leftarrow B$. 
Thus, it aligns with the observation in the above two sections
that metagraphs with style like $U\rightarrow *\leftarrow U\rightarrow B$ are better than those like $U\rightarrow B \rightarrow * \leftarrow B$. These similar observations described in these three sections can be regarded as domain knowledge, which indicates that we should design more metagraphs with style $U\rightarrow *\leftarrow U\rightarrow B$.

Finally, for $M_9$ on Yelp-50K and $M_6$ on Amazon-50K, we can see that their performance are among the best three, which demonstrates the usefulness of the complex semantics captured in $M_9$ on Yelp-50K and $M_6$ on Amazon-50K.

\subsection{Analysis of Other Components of FMG}
\label{sec-aspect-fmg}
We conduct experiments to further analyze other important components of the proposed ``MF + FM'' framework in Figure~\ref{fig-framework}, including the feature extraction methods in the MF part, the influence of the rank of second-order weight matrix in FM part, and comparisons between different optimization algorithms in the FM part.

\subsubsection{Feature Extraction Methods}
\label{sec-exp-mf-vs-nn}

\begin{figure}[ht]
	\centering
	\subfigure[Amazon-50K.]{%
		\label{fig-amazon-nn-vs-mf}
		\includegraphics[width=0.40\textwidth]{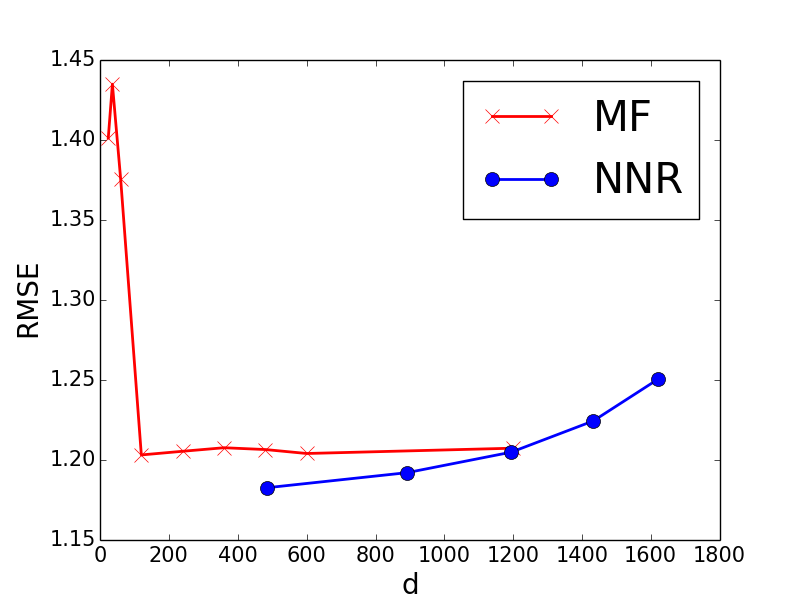}
	}
	\subfigure[Yelp-50K.]{
		\label{fig-yelp-nn-vs-mf}
		\includegraphics[width=0.40\textwidth]{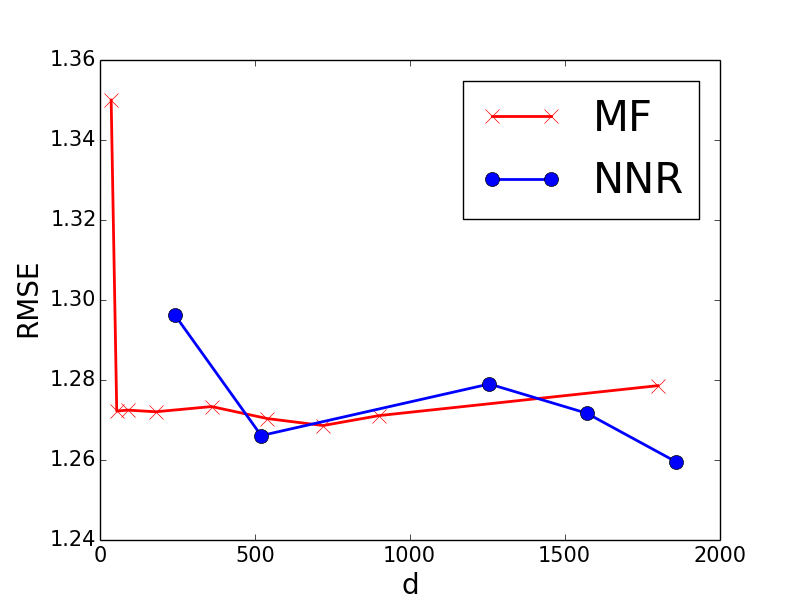}
	}
	\vspace{-10px}
	\caption{The performance of latent features obtained from MF and NNR.}
	\label{fig-nn-vs-mf}
\end{figure}

In this part, we compare the performance of different feature extraction methods in MF part, i.e., NNR and MF described in Section~\ref{sec-mg-sim}.
Note that, the parameter $F$ of MF and $\mu$ of NNR will lead to different number of latent features for different similarity matrices. 
Figure~\ref{fig-nn-vs-mf} shows the performance with different $d$,
i.e., total length of the input features.
We can see that latent features from NNR have slightly better performance than MF,
while the feature dimension resulting from NNR is much larger. 
These observations support our motivation to use these two methods in Section~\ref{sec-mg-latent-features}, that is, NNR usually has better performance while the recovered rank is often much higher than that of MF.
Thus, we can conclude that, 
NNR is better for extracting features, while MF is more suitable for a trade-off between performance and efficiency.

\subsubsection{Rank of Second-Order Weights Matrix}
\label{sec-exp-vary-K}

\begin{figure}[ht]
	\centering
	\includegraphics[width=0.40\textwidth]{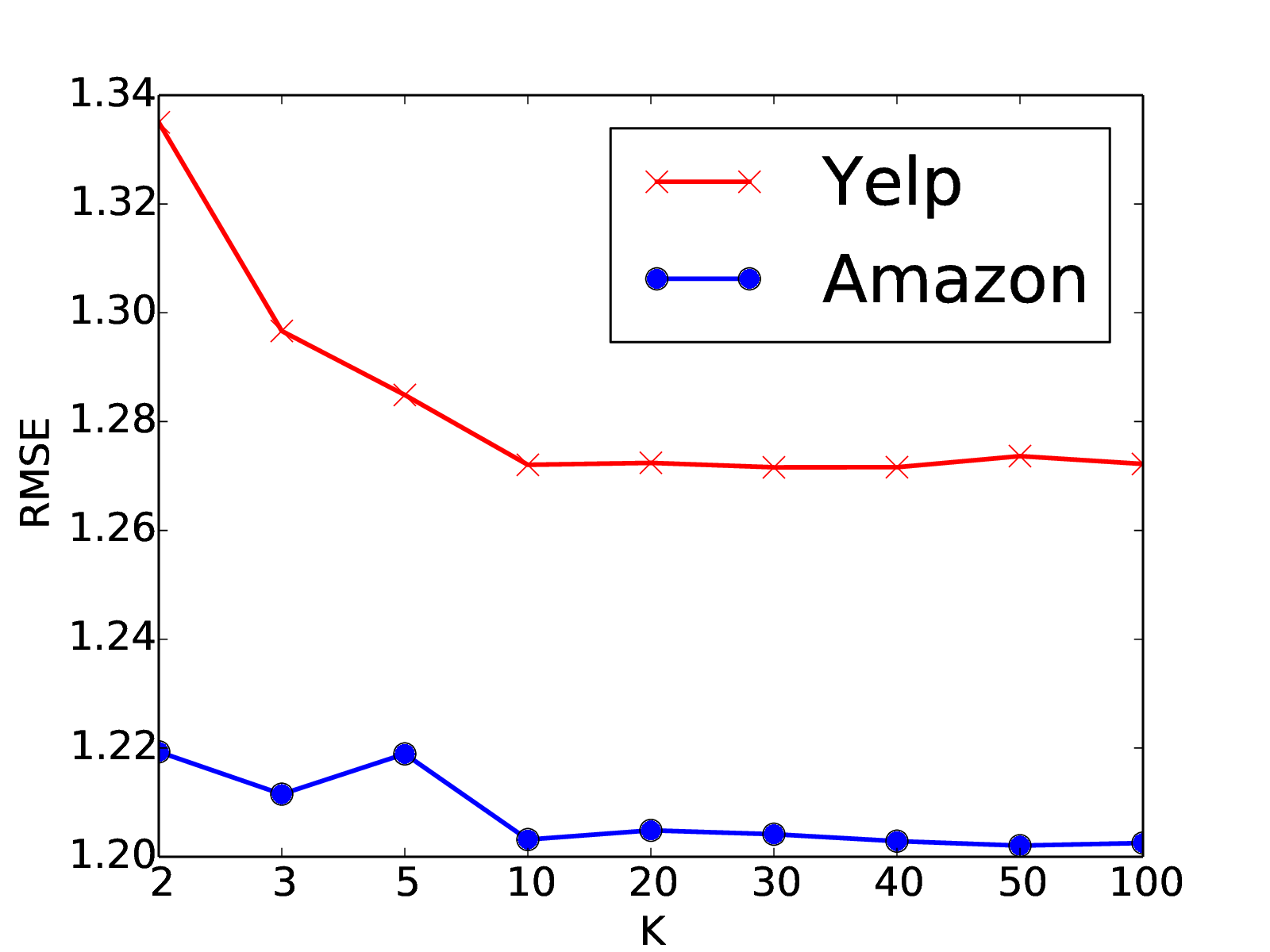}
	\vspace{-10px}
	\caption{The trend of RMSE of FMG w.r.t. $K$.}
	\label{fig-vary-K}
\end{figure}

In this part, we show the performance trend by varying $K$, 
which is the rank of the second-order weights $\textbf{V}$ in the FMG model (see Section~\ref{sec-framework-model}). 
For the sake of efficiency, we conduct extensive experiments on Amazon-50K and Yelp-50K and employ the MF-based latent features. We set $K$ to values in the range of $[2,3,5,10,20,30,40,50,100]$, and the results are shown in Figure~\ref{fig-vary-K}.
We can see that the performance becomes better with larger $K$ values on both datasets and reaches a stable performance after $K = 10$.
Thus, we fix $K = 10$ for all other experiments.

\subsubsection{Optimization Algorithm}
\label{sec-exp-svrg-vs-nmAPG}
Here, 
we compare the SVRG and nmAPG algorithms proposed in Section~\ref{sec:opt}. Besides, we also use SGD as a baseline since it is the most popular algorithm for models based on factorization machine \cite{rendle2012fm,hong2013co}.
Again, we use the Amazon-50K and Yelp-50K datasets. As suggested in \cite{xiao2014proximal}, we compare the efficiency of various algorithms based on RMSE w.r.t. the number of gradient computations divided by $N$.

\begin{figure}[ht]
	\centering
	\subfigure[FMG@Amazon-50K.]
	{\includegraphics[width=0.40\textwidth]{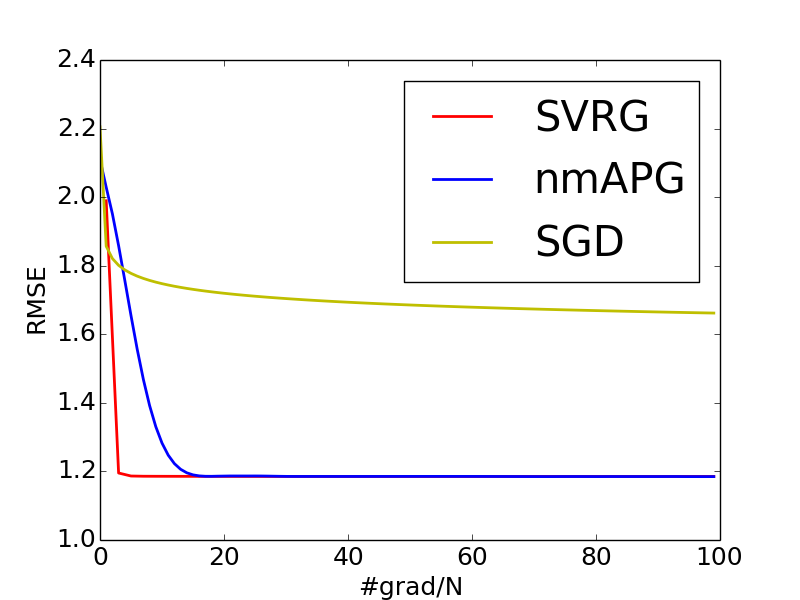}}
	\subfigure[FMG(LSP)@Amazon-50K.]
	{\includegraphics[width=0.40\textwidth]{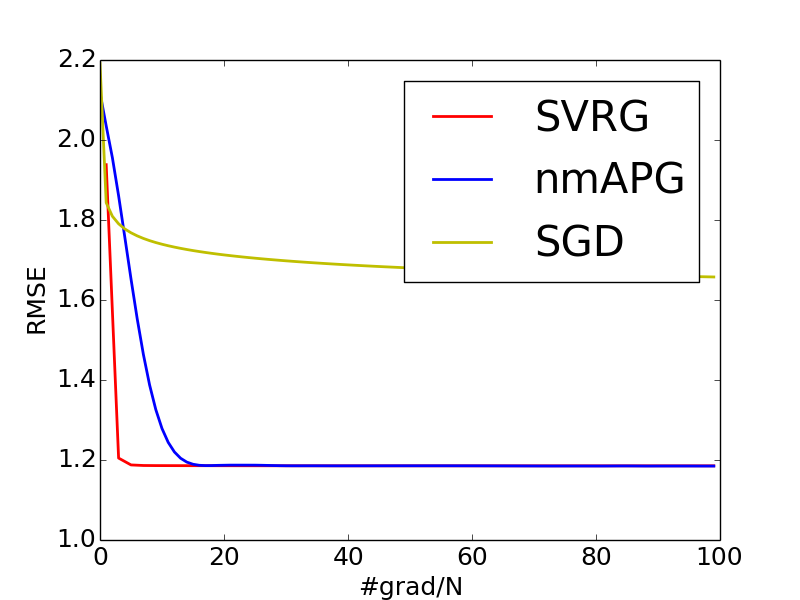}}
	
	\vspace{-10px}
	
	\subfigure[FMG@Yelp-50K.]
	{\label{fig-2}
		\includegraphics[width=0.40\textwidth]{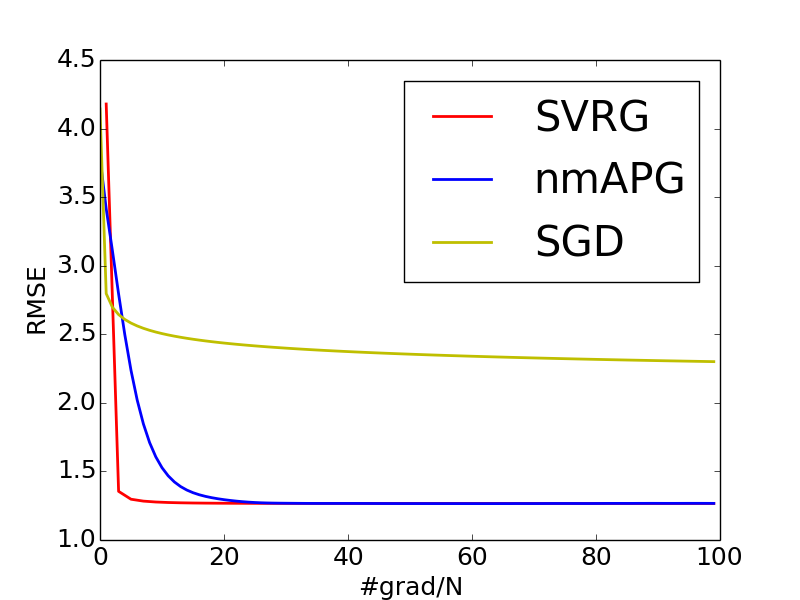}}
	\subfigure[FMG(LSP)@Yelp-50K.]
	{\label{fig-4}
		\includegraphics[width=0.40\textwidth]{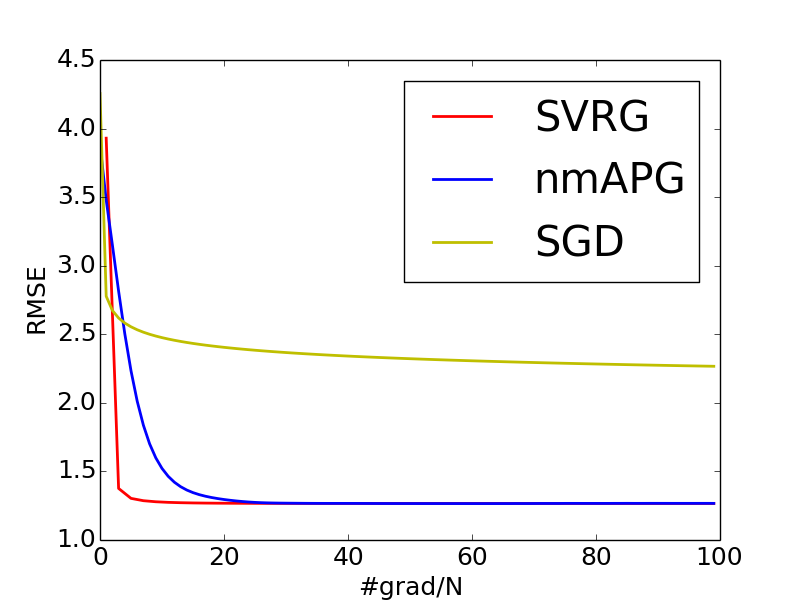}}
	\vspace{-10px}
	\caption{Comparison of various algorithms on the Amazon-50K and Yelp-50K datasets.}
	\label{fig:obj}
\end{figure}

The results are shown in Figure~\ref{fig:obj}.
We can observe that SGD is the slowest among all three algorithms and SVRG is the fastest. Although SGD can be faster than nmAPG at the beginning, the diminishing step size used to guarantee convergence of stochastic algorithms finally drags SGD down to become the slowest. SVRG is also a stochastic gradient method, but it avoids the problem of diminishing step size using variance reduced technique, which results in even faster speed than nmAPG.
Finally, as both SVRG and nmAPG are guaranteed to produce a critical point of \eqref{eq-obj}, they have the same empirical prediction performance. Therefore, in practice, the suggestion is to use SVRG as the solver because of the faster speed and empirically good performance.

%
%
%

\subsection{Parameter Analysis of HAF}
\label{sec-exp-analysis-haf}
In this section, we conduct more experiments to show the influences of different parameters of the proposed HAF. The following three parameters are chosen: the output embedding size, the number of heads in node-level attention, and the metagraph-level attention vector size. We run the experiments with HAF by varying the corresponding parameters, and report the trend of test RMSEs. Finally, we visualize the attention weights to show the importance of each metagraph.

\subsubsection{The Influence of the Output Embedding Size}
\begin{figure}[ht]
	\centering
	\subfigure[Amazon.]
	{\includegraphics[width=0.40\textwidth]{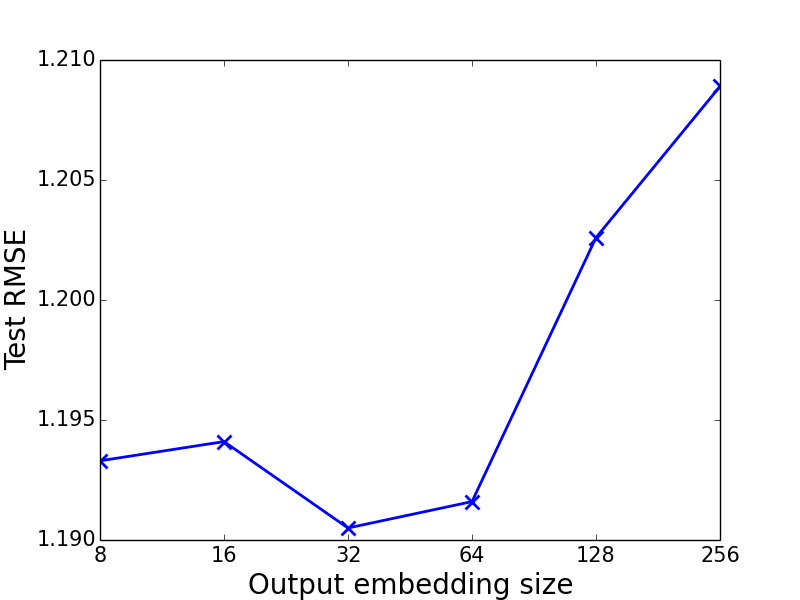}}
	\subfigure[Yelp.]
	{\includegraphics[width=0.40\textwidth]{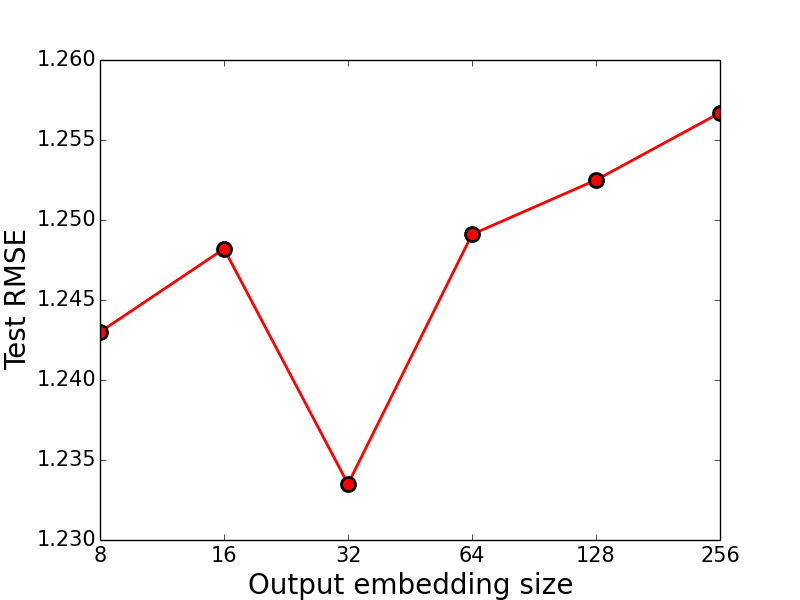}}
	\vspace{-10px}
	\caption{The trend of test RMSE w.r.t the output embedding size of HAF.}
	\label{fig-haf-output-size}
\end{figure}

The output embedding size is varied in $[8, 16, 32, 64, 128, 256]$, and the performance trending is shown in Figure \ref{fig-haf-output-size}. We can see that with the increase of the output embedding size, the RMSE decreases firstly and then increases, which means that either too large or too small sizes of the output embedding will impair the performance. Thus, we set it to $32$ in Section~\ref{sec-exp-rmse} according to Figure~\ref{fig-haf-output-size}. 

\begin{figure}[ht]
	\centering
	\subfigure[Amazon.]
	{\includegraphics[width=0.40\textwidth]{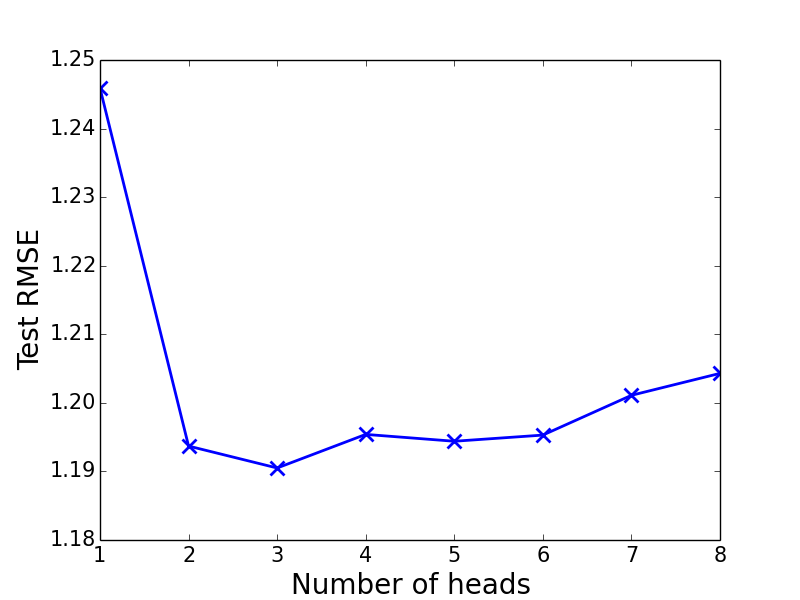}}
	\subfigure[Yelp.]
	{\includegraphics[width=0.40\textwidth]{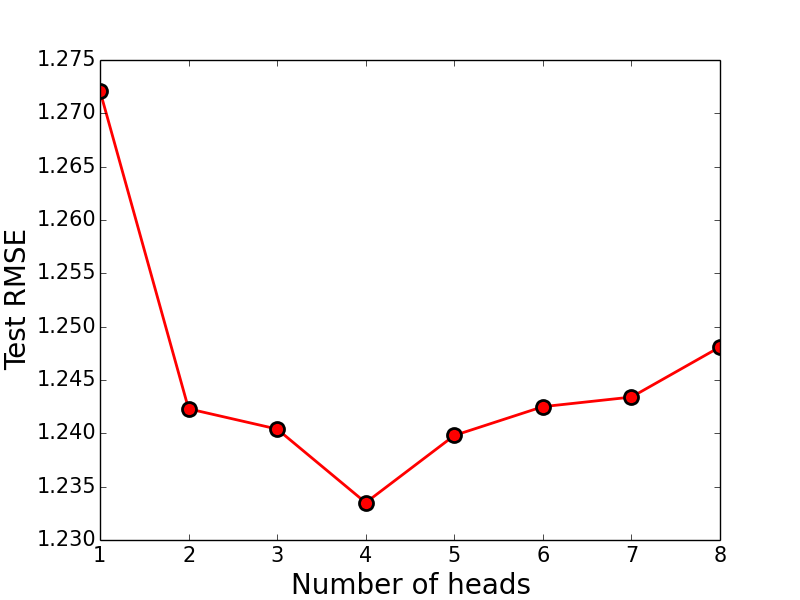}}
	\vspace{-10px}
	\caption{The trend of test RMSE w.r.t the number of heads in node-level attention of HAF.}
	\label{fig-haf-gat-head}
\end{figure}

\subsubsection{The Influence of the Number of Heads in Node-level Attention.}
As shown in Eq.~\eqref{eq-self-attention}, the number of heads can affect the output embedding of the node-level attention and thus the final performance. We vary the number of heads in $[1,2,3,4,5,6,7,8]$, and the performance trends are shown in Figure~\ref{fig-haf-gat-head}. We can see that HAF achieves the best performance when the number of heads is $3$ (Amazon) or $4$ (Yelp). It does not necessarily improve the performance when the number of heads is larger, which is the case in other works~\cite{wang2019heterogeneous}. 

\begin{figure}[ht]
	\centering
	\subfigure[Amazon.]
	{\includegraphics[width=0.40\textwidth]{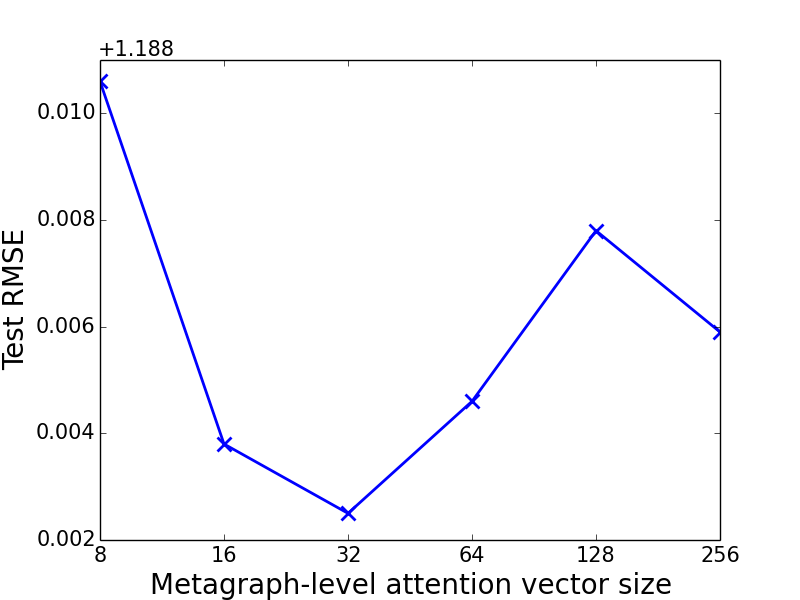}}
	\subfigure[Yelp.]
	{\includegraphics[width=0.40\textwidth]{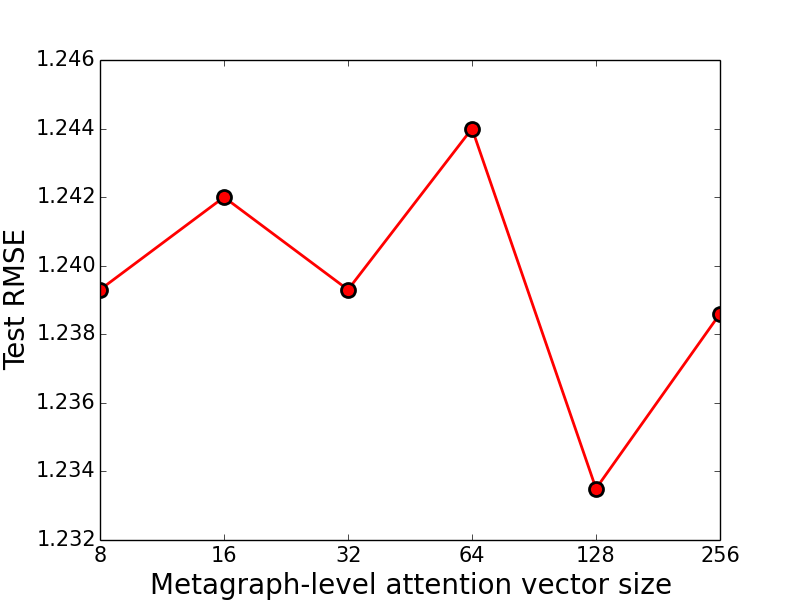}}
	\vspace{-10px}
	\caption{The trend of test RMSE w.r.t the metagraph-level attention vector size of HAF.}
	\label{fig-haf-sem-att-size}
\end{figure}

\subsubsection{The Influence of the Metagraph-level Attention Vector Size}
The metagraph-level attention size is denoted by $\bq$ in Eq.~\eqref{eq-metagraph-attention}, which affects the fusion of metagraph based embeddings. The size is varied in $[8, 16, 32, 64, 128, 256]$, and the results are shown in Figure \ref{fig-haf-sem-att-size}. We can see that the trend is similar to that of the output embedding size, i.e., a moderate embedding size is important for the final performance.

\begin{figure}[H]
	\centering
	\subfigure[Amazon.]
	{\includegraphics[width=0.40\textwidth]{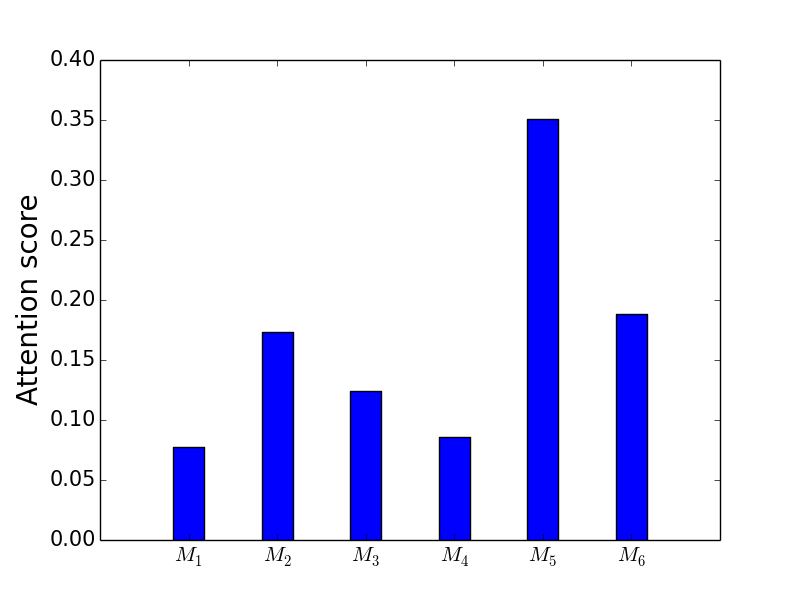}}
	\subfigure[Yelp.]
	{\includegraphics[width=0.40\textwidth]{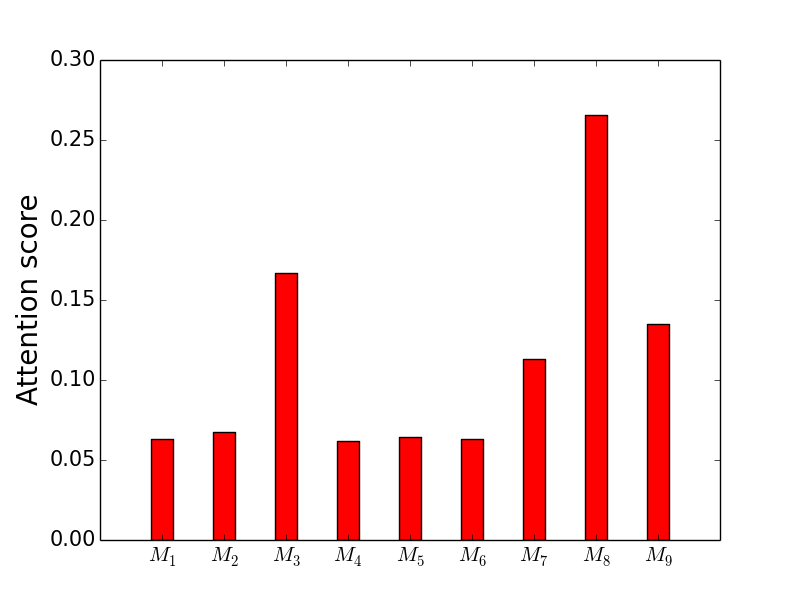}}
	\vspace{-10px}
	\caption{The attention distribution over all metagraphs in HAF.}
	\label{fig-haf-att-dis}
\end{figure}
\subsubsection{The Attention Weight Distribution}
We further visualize the attention weights of different metagraphs, which are shown in Figure~\ref{fig-haf-att-dis}. We can see that on Amazon, $M_5$ is the most important, and on Yelp, $M_8$ is the most important. When looking at Figure~\ref{fig-amazon-metagraph} and \ref{fig-yelp-metagraph}, both of two metagraphs are $U\rightarrow R \rightarrow A \rightarrow U\rightarrow B$, which corresponds to the aspects of textual information. It might be the reason that on websites like Amazon and Yelp, the textual content plays a very important role in users' preferences and thus the recommendation performance. The metagraph-level attention can capture this pattern from the data. 

\section{Conclusion and Future Works}
\label{sec-conclusion}

In this paper, we address the side information fusion problem in CF based recommendation scenarios over heterogeneous information networks (HIN).
By using metagraphs derived given a HIN, we can capture similarities of complex semantics between users and items, based on which two fusion frameworks are proposed.
The first one is the ``MF + FM'' framework. From each metagraph, we obtain a user-item matrix, to which we apply matrix factorization and nuclear norm regularization to obtain the user and item latent features in an unsupervised way. After that, we use a group lasso regularized factorization machine to fuse different groups of latent features extracted from different metagraphs to predict the links between users and items, i.e., to recommend items to users. 
The second is a deep learning model, the hierarchical attention fusion (HAF) framework, which tries to extract and fuse metagraph based latent features in an end-to-end manner.
Extensive experimental results demonstrate the effectiveness of the two proposed frameworks.

In the future, we plan to explore automatic methods~\cite{quanming2018taking,zhao2020simplifying,ding2020propagation} to generate metagraphs instead of hand-crafting them as done in this paper. Thus, our frameworks can be quickly applied to new domains.

\section{Acknowledgments}
Huan Zhao and Dik Lun Lee are supported by the Research Grants Council HKSAR GRF (No. 16215019). Quanming Yao and James T. Kwok are supported by the Research Grants Council HKSAR GRF (No. 614513).
Yangqiu Song is supported NSFC (U20B2053), Hong Kong RGC including Early Career Scheme (ECS, No. 26206717), General Research Fund (GRF, No. 16211520), and Research Impact Fund (RIF, No. R6020-19), and WeBank-HKUST Joint Lab.
We also thank the anonymous reviewers for their valuable comments and suggestions that help improve the quality of this manuscript.

\bibliographystyle{ACM-Reference-Format}
\bibliography{main} 

\end{document}